\documentclass[usegraphicx]{mn2e}
\usepackage{amsmath,amsfonts,amsopn,amssymb,stmaryrd,xspace}
\usepackage{tablefootnote}

\newcommand{\most}{\textit{MOST}\xspace}

\title[The Massive, Interacting Binary MWC 314]{Spectroscopy, \most Photometry, and Interferometry of MWC 314: Is it an LBV or an interacting binary?}

\author[N. D. Richardson et al.]{Noel~D.~Richardson$^{1}$\thanks{E-mail:richardson@astro.umontreal.ca},
Anthony F. J. Moffat$^{1}$,
Rapha{\"e}l Maltais--Tariant$^{1}$,
\newauthor Herbert Pablo$^{1}$,
Douglas R. Gies$^{2}$, 
Hideyuki Saio$^{3}$,
Nicole St-Louis$^{1}$,
Gail Schaefer$^{4}$,
\newauthor Anatoly S. Miroshnichenko$^{5}$,
Chris Farrington$^{4}$,
Emily J. Aldoretta$^{1}$,
\newauthor {\'E}tienne Artigau$^{1}$,   
Tabetha S. Boyajian$^{6}$, 
Kathryn Gordon$^{2}$, 
Jeremy Jones$^{2}$, 
\newauthor Rachel Matson$^{2}$,
Harold A. McAlister$^{2}$, 
David O'Brien$^{7}$, 
Deepak Raghavan$^{2}$, 
\newauthor Tahina Ramiaramanantsoa$^{1}$,
Stephen T. Ridgway$^{8}$,
Nic Scott$^{4}$,
Judit Sturmann$^{4}$,
\newauthor Laszlo Sturmann$^{4}$,
Theo ten Brummelaar$^{4}$,
Joshua D. Thomas$^{9}$,
Nils Turner$^{4}$, 
\newauthor Norm Vargas$^{4}$, 
Sergey Zharikov$^{10}$,
Jaymie Matthews$^{11}$, 
Chris Cameron$^{12}$, 
\newauthor David Guenther$^{13}$, 
Rainer Kuschnig$^{11,14}$, 
Jason Rowe$^{15}$, 
Slavek Rucinski$^{16}$, 
\newauthor Dimitar Sasselov$^{17}$, 
and Werner Weiss$^{14}$ \\
$^{1}$ D\'epartement de physique and Centre de Recherche en Astrophysique du Qu\'ebec (CRAQ), Universit\'e de Montr\'eal, C.P. 6128,\\  Succ.~Centre-Ville, Montr\'eal, Qu\'ebec, H3C 3J7, Canada\\
$^{2}$ Center for High Angular Resolution Astronomy, Department of Physics and Astronomy, Georgia State University, P. O. Box 5060,\\  Atlanta, GA 30302-5060, USA \\
$^{3}$ Astronomical Institute, Graduate School of Science, Tohoku University, Sendai, Miyagi 980-8578, Japan\\
$^{4}$ The CHARA Array, Mount Wilson Observatory, 91023 Mount Wilson CA, USA \\
$^{5}$ Department of Physics and Astronomy, University of North Carolina at Greensboro, Greensboro, NC 27402-6170, USA \\
$^{6}$ Yale University, New Haven, CT 06520-8101, USA \\
$^{7}$ Max Planck Institute for Radio Astronomy, P.O.~Box 20 24, D-53010 Bonn, Germany\\
$^{8}$ National Optical Astronomy Observatory, 950 North Cherry Ave., Tucson, AZ 85719, USA \\
$^{9}$ Department of Physics, Clarkson University, 8 Clarkson Ave, Potsdam, New York 13699, USA \\
$^{10}$ Instituto de Astronom\'{i}a, Universidad Nacional Aut\'{o}noma de M\'{e}xico, Ensenada, BC 22860, Mexico \\
$^{11}$ {Department of Physics and Astronomy, University of British Columbia, 6224 Agricultural Road, Vancouver, BC V6T 1Z1, Canada}\\
$^{12}$ {Department of Mathematics, Physics \& Geology, Cape Breton University, 1250 Grand Lake Road, Sydney, Nova Scotia B1P 6L2, Canada}\\
$^{13}$ {Institute for Computational Astrophysics, Dept. of Astronomy and Physics, St Mary's University Halifax, NS B3H 3C3, Canada}\\
$^{14}$ {University of Vienna, Institute for Astronomy, T\"urkenschanzstrasse 17, A-1180 Vienna, Austria}\\
$^{15}$ {NASA Ames Research Center, Moffett Field, CA 94035, USA}\\
$^{16}$ {Dept. of Astronomy and Astrophysics, University of Toronto, 50 St George Street, Toronto, ON M5S 3H4, Canada}\\
$^{17}$ {Harvard-Smithsonian Center for Astrophysics, 60 Garden Street, Cambridge, MA 02138, USA}\\
}

\begin{document}
\bibliographystyle{mn2e}

\date{}

\pagerange{\pageref{firstpage}--\pageref{lastpage}} \pubyear{2015}

\maketitle

\label{firstpage}

\clearpage
\begin{abstract}

MWC 314 is a bright candidate luminous blue variable that resides in a fairly close binary system, with an orbital period of $60.753\pm0.003$ d. We observed MWC 314 with a combination of optical spectroscopy, broad-band ground- and space-based photometry, as well as with long baseline, near-infrared interferometry. We have revised the single-lined spectroscopic orbit and explored the photometric variability.
The orbital light curve displays two minima each orbit that can be
partially explained in terms of the tidal distortion of the primary
that occurs around the time of periastron.
The emission lines in the system are often double-peaked and stationary in their kinematics, indicative of a circumbinary disc. 
We find that the stellar wind or circumbinary disc is partially resolved in the $K'$-band with the longest baselines of the CHARA Array. 
From this analysis, we provide a simple, qualitative model in an attempt to explain the observations. From the assumption of Roche Lobe overflow and tidal synchronisation at periastron, we estimate the component masses to be $M_{1} \approx 5 M_{\odot}$ and $M_{2} \approx 15 M_{\odot}$, which indicates a mass of the LBV that is extremely low. In addition to the orbital modulation, we discovered two pulsational modes with the \most satellite. These modes are easily supported by a low-mass hydrogen-poor star, but cannot be easily supported by a star with the parameters of an LBV.
The combination of these results provides evidence that the primary star was likely never a normal LBV, but rather is the product of binary interactions. As such, this system presents opportunities for studying mass-transfer and binary evolution with many observational techniques.

\end{abstract}


\section{Introduction}

Massive stars provide much of the energy input in the Universe. Their high mass-loss rates and supernova explosions provide important feedback to the star formation processes and total energy input of galaxies. Recent advances in stellar modelling (e.g., Groh et al.~2013) show that the supernova progenitors for core collapse supernovae come in many different types that include red supergiants (RSG), blue supergiants (BSG), Wolf-Rayet (WR) stars, and Luminous Blue Variables (LBVs). 

Massive stars tend to be found primarily in binary systems. The O stars are thought to have a bound companion about 75\% of the time, with most of the exceptions being runaway stars (e.g. Mason et al.~2009). Mason et al.~found that O stars in clusters and associations have companions at least 60-80\% of the time. Their sample focused on high angular resolution techniques, but also incorporated spectroscopic results. Sana et al.~(2012) showed that 71\% of O stars will have a binary interaction during their lives. They found that only 29\% of the O stars are effectively single (either very-long period binary stars or actually single), meaning that evolutionary models that do not incorporate binary effects will have limited applicability. 

LBVs are among the most unusual classes of massive stars. They have attained a highly luminous, unstable state that shows remarkable mass-loss and variability. The normal mass-loss rates range between $10^{-6}-10^{-3}$ M$_{\odot}$ yr$^{-1}$, which has typically led to the conclusion that these objects are post-main sequence, hydrogen-shell-burning massive stars that represent the transitionary phase between the main sequence O stars and the helium-burning WR stars (e.g.~Humphreys \& Davidson 1994; van Genderen 2001). However, the recent analysis by Groh et al.~(2013) shows that lower initial mass stars (20--25 $M_\odot$) can become LBVs after the RSG phase, and then explode as type II supernovae during the final LBV phase.

With a large binary fraction for the main-sequence O stars, one may also expect to see a high binary fraction in the post-main sequence massive stars. However, the binary fraction for WR stars is low (40\%) as noted by Vanbeveren \& Conti (1980). An examination of the multiplicity and binarity of LBVs was reported by Martayan et al. (2012) who reported a remarkably low binary fraction of only 11\%. There are only four well-studied LBV binary systems: $\eta$ Carinae (e.g.~Richardson et al.~2010, Madura et al.~2013 and references therein), MWC 314 (Lobel et al.~2013, hereafter L13), HD 5980 (e.g.~Koenigsberger et al.~2010), and R 81 (HDE 269128; Tubbesing et al.~2002). $\eta$ Car is by far the most studied of all these binary systems, but analyses of both $\eta$ Car and the other systems have not produced reliable, model-independent masses yet. For example, in the R 81 system, a clear eclipse is seen in the light curve while the primary is in front of the secondary in our line of sight, but the secondary eclipse is small, and similar in amplitude to the pulsations in the system. The masses of both stars are only known to a factor of $\sim 2$, which does not provide much insight into the masses of these evolved massive stars (Tubbesing et al.~2002). HD 5980 shows orbital variations due to both colliding winds and binarity, and the system shows long-term evolution in its light curve similar to that of other LBVs (Koenigsberger et al.~2010), but the spectrum reveals that the two stars both appear as WNh stars (Wolf-Rayet stars showing nitrogen-enrichment and hydrogen). This makes the temperature of HD 5980 hotter than all other LBVs so a determination of its mass may not be typical of LBVs.
MWC 314 is unique in that the orbital period is semi-short (60.8 d), it is well-placed in the sky ($\delta = +14^{\circ}$), and bright enough to allow high resolution studies of the star and its environs with multiple observing techniques. 

MWC 314 (BD+14 3887, V1429 Aql, Hen 3-1745) has been examined in a few key studies, which were largely reviewed in the recent work of L13. Miroshnichenko (1996) found the star to be of high luminosity, exhibit a strong interstellar reddening, and to have a similar luminosity to the prototypical LBV, P Cygni. Miroshnichenko et al.~(1998) then determined the spectral characteristics and found it to be very similar to P Cygni, with an estimated distance of $3.0\pm0.2$ kpc.
Wisniewski et al.~(2006) presented a long-term spectropolarimetric and spectroscopic study of MWC 314, finding the first evidence of H$\alpha$ variability. They found that the polarisation was variable in a way reminiscent of an asymmetric wind. 
MWC 314 was shown to be a spectroscopic binary by Muratorio et al.~(2008), but they underestimated the orbital period. L13 measured absorption lines of S II and Ne I, which are thought to be photospheric, and found the system to have a 60.8 d orbit with a small eccentricity of $e=0.23$. They also demonstrated that the $V-$band photometry from the All Sky Automated Survey is modulated on the orbital time scale, and their interpretation of the light curve includes two partial eclipses.

L13~used a model of the single-lined spectroscopic radial velocity orbit and photometric light curve to help estimate the mass ratio and system parameters. They found that the primary star had typical parameters for an LBV, with $T_{\rm eff}=18,000$K, $\log g = 2.26$, $M \simeq 40 M_\odot$, and $R \simeq 87 R_\odot$. The secondary star's parameters were very unusual in that they suggest the companion is a yellow giant. This interpretation is inconsistent from an evolutionary standpoint because the secondary should not be able to reach an advanced evolutionary state (cool temperature) given the short lifetime of the evolved primary star. Liermann et al.~(2014) present NIR $K-$band spectrophotometry of MWC 314 and some B[e] stars. Several of these stars show CO spectral features that would be consistent with a companion similar to that suggested by L13, but MWC 314 does not show these features. 

L13~also developed a three-dimensional wind model to create synthetic He I wind lines for comparative purposes showing some evidence for an asymmetric wind, which was further developed by Lobel et al.~(2015). This asymmetric wind has a density enhancement on the leading hemisphere of the LBV that feeds gas into a circumbinary disc.
Further, they obtained an image of the H$\alpha$ emission nebula surrounding the star which shows a bipolar structure on large scales (Marston \& McCollum~2008), but appears spherically symmetric at small scales of a few arcseconds.

In this paper, we present a variety of new observations (spectroscopy, photometry, and long-baseline near-infrared interferometry) of the MWC 314 system, which is described in Section 2. 
In Section 3, a revised single-lined orbit based upon new spectroscopy and the work of L13 is presented and discussed. 
Section 4 discusses the orbital light-curve, as well as the discovery of pulsational modes. 
Our interferometric results are presented in Section 5. 
In Section 6, we present a general discussion of the system with respect to the fundamental parameters, pulsations, and the interferometric results. 
We conclude our study in Section 7. 
 
\section{Observations}
\subsection{Spectroscopy}

We observed MWC 314 with a variety of telescopes and instruments with the primary goal being to better constrain the single-lined orbit. The telescopes used include the CTIO 1.5 m operated by the SMARTS Consortium, the Observatoire du Mont M\'{e}gantic 1.6 m, McDonald Observatory's Struve 2.1 m and Harlan Smith 2.7 m, the Mercator 1.2 m, and the San Pedro Martir 2.1 m telescopes. All observations were reduced using standard techniques for long-slit or echelle spectroscopy utilising bias frames and flat fields with either IRAF\footnote{IRAF is distributed by the National Optical Astronomy Observatory, which is operated by the Association of Universities for Research in Astronomy (AURA) under cooperative agreement with the National Science Foundation.} or custom software. Wavelength calibration was accomplished through emission-line comparison spectra taken before or after each observation.
A spectroscopic observing log is presented in Table A1 that details the telescopes, spectrographs, and data.

L13 measured the absorption lines S II $\lambda\lambda 5454, 5474, 5647$ and Ne I $\lambda 6402$, due to their lack of blending with the large number of emission lines in the spectrum of MWC 314. We sought to include these lines whenever possible. Radial velocity measurements were made through Gaussian fits of the spectral lines. We found that we were able to use the Ne I $\lambda 6402$ line for all data sets, but the S II lines often suffered from lower S/N and were unreliable. The McDonald Observatory observations provided excellent data for S II $\lambda 5647$, but were not usable for the S II $\lambda\lambda 5454, 5474$ lines. The data from the CTIO 1.5 SMARTS fiber-fed bench-mounted echelle\footnote{http://www.ctio.noao.edu/$\sim $atokovin/echelle/FECH-overview.html} (Barden \& Ingerson 1998) were only usable around the Ne I line. 
However, the data obtained with the CTIO 1.5 m and the CHIRON spectrograph (Tokovinin et al.~2013) have higher S/N across the optical spectrum.
The spectroscopic data discussed in L13 were obtained with the Mercator 1.2 m telescope and the HERMES spectrograph (Raskin et al.~2011), which we also used to obtain three additional spectra of comparable quality.
The spectrum taken at San Pedro Martir was extremely useful due to the orbital phase observed (most negative radial velocity) and good S/N.
The data from the Observatoire du Mont M\'{e}gantic (OMM) were of much lower spectral resolution, and we only obtained two spectra that had high enough S/N to measure accurately a radial velocity from the weak absorption lines. 

\subsection{Ground-Based Photometry}

Broadband $BVRI$ photometry was obtained throughout the 2010--2012 calendar years with the American Association of Variable Star Observers automated telescope located at Lowell Observatory. The telescope\footnote{http://www.aavso.org/w28} is a Celestron C11 Schmidt-Cassegrain instrument with an aperture of 28 cm. Imaging was accomplished with an SBIG ST-7XME camera, which yields images with a field of view of 14$\times$9 arcmin. The reductions utilise dark, bias, and flat frames, and typically the scattered light background leads to a photometric accuracy of 1--2\%. The differential photometry was done relative to another star in the field, HD 231237, which shows a constant light curve in the All Sky Automated Survey (ASAS; Pojma\'{n}ski 2002) and was classified as G5 by Cannon (1925). 

We wished to obtain a reasonable magnitude estimate of MWC 314 relative to the comparison star, so magnitudes for the comparison star were obtained from the AAVSO Photometric All-Sky Survey (APASS)\footnote{http://www.aavso.org/apass}, which measured $B = 11.48 \pm 0.047$, $V = 10.716 \pm 0.027$, $g' = 11.024 \pm 0.032$, $r' = 10.497 \pm 0.016$, and $i' = 10.228 \pm 0.054$ for HD 231237. The Johnson $R$ magnitude was obtained by using the relations given by Kent (1985), which gave $R = 9.95$. Similarly, we transformed these data into Johnson $I$ using the transformations given by Windhorst et al.~(1991), which yields $I = 9.48$. We present the measurements of MWC 314 calibrated by HD 231237 in Table A4. However, we caution that we did not perform any colour transformations on the data, so the magnitudes may have small systematic errors related to airmass at the time of observation.

\subsection{Precision Photometry from \most}

We observed the system with the micro-satellite \most ({\it Microvariability and Oscillations of STars}) which has a 15-cm Maksutov telescope with a custom broad-band filter covering 3500--7500 \AA. The sun-synchronous polar orbit has a period of 101.4 minutes ($f=14.20 \ {\rm d}^{-1}$), which enables uninterrupted observations for up to eight weeks for targets in the continuous viewing zone. A pre-launch summary of the mission is given by Walker et al.~(2003). The satellite was never intended to observe a target for several months and recover time-scales on the order of the length of the data. 

MWC 314 was observed for a small portion of every spacecraft orbit for 55 d, spanning 2014 June 19 to 2014 August 15 in the direct imaging mode. Our data were taken over short orbital segments, which we then averaged during each 5--10 minute interval to have precise photometry from the orbital means. The photometry was extracted using the standard \most pipeline (Walker et al.~2003), and we show two different versions of the final light-curve in this paper, which are given in Tables A6 and A7. 
The first uses a trend-removed data that removes the binary-induced signal from the light curve. We also attempted to extract the binary light curve by using a raw extraction that allowed for the signal to remain. 
It was difficult to recover the binary signal, as a remaining instrumental response needed to be removed through a comparison with all guide stars that were observed simultaneously, and then fitting and removing an average ``instrumental" trend from the data. We note that a small portion of this light curve could not be corrected for the instrumental response. 
The long time-series from \most was not continuous due to data gaps induced from passages through the South Atlantic Anomaly, problems with scattered light that is more prominent during northern summer months, and communications errors. Nevertheless, the {\it MOST} observations provide a unique photometric data set to explore the variability of this object.

\subsection{Long Baseline, Near-Infrared Interferometry}

We obtained multiple epochs of long baseline near-infrared interferometry using the CHARA Array and the Classic (ten Brummelaar et al. 2005) and CLIMB beam combiners (ten Brummelaar et al.~2013) in the $K'-$band during the calendar years 2010--2013. The CHARA Array is a $\Ydown$-shaped interferometric array of six 1-m telescopes with baselines ranging from 34 to 331 meters in length. Our observations were primarily at longer baselines, but we obtained a few measurements with short baselines. The nights of observations, baselines used, and calibrators for each observation are listed in Table A2.

To measure the instrument response and calibrate our data, we observed calibrator stars with small angular diameters both before and after each observation of MWC 314. Namely, we observed the calibrator stars 
HD 174897 ($\theta_{LD} = 0.652 \pm 0.038$ mas; Boyajian et al.~2012), 
HD 182101 ($\theta_{LD} = 0.367 \pm 0.017$ mas; Berger et al.~2006, Baines et al.~2010), and
HD 184606 ($\theta_{LD} = 0.236 \pm 0.050$ mas; van Belle et al.~2008). 
These calibrators have diameters known from fits to the spectral energy distribution and have all been reliable for previous interferometric studies.
The data were reduced using the standard CHARA reduction pipeline (ten Brummelaar et al.\ 2005, 2013).  The visibilities and closure phases were averaged over each observing block. The calibrated OIFITS data files (Pauls et al. 2005) will be available through the JMMC archive\footnote{http://www.jmmc.fr/oidb.htm} or upon request, but we also include tabulated visibilities in Table A5.

Schaefer et al.~(2014) describe the effects of emission lines on the errors and measurements of $V^2$. An emission line reduces the effective bandpass over which the fringe amplitude is measured causing the true visibility to be smaller than if a fixed bandpass was assumed. We measured the effective bandpass through comparisons of the width of the power spectra of both calibrator stars and MWC 314, and found that this is typically an effect of $<3\%,$ much smaller than the typical error of our measurements. Further, we compared a $K'-$band filter response with a single NIR spectrum of MWC 314 we obtained with the Mimir instrument and the Lowell Observatory 1.8 m telescope (Clemens et al.~2007). This spectrum was reduced with the standard Mimir pipline\footnote{http://people.bu.edu/clemens/mimir/software.html} and flux-calibrated and telluric-corrected using the {\tt xtellcor} package (Vacca et al.~2003). This spectrum was then convolved with the filter response, and we found that the effective width was within 1-2\% of the nominal width. These values show that there was little change in the visibilities, but the resulting uncertainty in the width of the filter ($\sim 3\%$) was added in quadrature to the pipeline-produced error.

\section{The Single-Lined Spectroscopic Orbit}

The optical spectrum of MWC 314 consists of four different kinds of spectral lines, which are illustrated in the spectral atlas of Miroshnichenko et al.~(1998). First, there are absorption lines, such as those measured by L13. These are thought to be photospheric in origin, with minimal contamination by the wind. These lines are similar to the photospheric metal lines observed in the B7 Ia supergiant HD 183143 (Chentsov et al.~2003), but are typical of any early- or mid-B supergiant. Secondly, the wind of the system is best seen in the strong Balmer lines and He I emission lines. The Balmer lines do not show a P Cygni absorption component, but the He I lines have P Cygni absorption that shows orbitally-modulated variability where the absorption strengthens at certain orbital epochs (L13). Thirdly, there are many lines of singly-ionised metals, such as Fe II, that show double-peaked emission profiles. These lines were shown by L13 to be constant in radial velocity and are thought to originate from a circumbinary disc. In reality, some of the double-peaked emission lines and Balmer lines are likely formed in a combination of the circumbinary disc and stellar wind. The fourth and final type of lines are the interstellar absorption lines. These lines are very complicated, often showing several absorption components due to the extreme extinction of MWC 314 ($A_V = 5.7$; Miroshnichenko 1996). 

We measured radial velocities through Gaussian fits of the same absorption lines as measured by L13 to refine the spectroscopic orbit of MWC 314. The lines used were discussed in Section 2.1, and the new radial velocities are presented in Table A3. We estimate the errors on most of the data points to be on the order of $\pm2$ km s$^{-1}$, and a little better ($\pm 1$ km s$^{-1}$) for the HERMES spectra. The data are shown in Figure 1 with our orbital elements presented in Table 1. The data span the calendar years 2001--2013, so we derived the orbital period through a time-series analysis of these new radial velocities combined with the velocities reported by both Muratorio et al.~(2008) and L13. We confirm the orbital period,
although we adopt a more conservative error than reported
by L13. Note that L13~set 
phase zero at inferior conjunction of the visible star
and also measure $\omega$ from this epoch, instead of 
the standard spectroscopic method of setting $\omega$ 
equal to the angle between the ascending node and periastron.
The values of $T$ and $\omega$ given in Table 1 for the 
L13 solution are referenced to periastron
in the standard way.  The errors reported by L13~are
generally smaller than ours, however, the PHOEBE code they used 
does not directly estimate the errors of the parameters, so we suspect that these are underestimated.  
We show orbital parameters in Table 1 for L13's fit, our fit to their data showing more realistic errors, an independent fit to only our new data, and a combined fit to all data available from Muratorio et al.~(2008), L13, and our new data. All the solutions are comparable, and we adopt the combined fit for the remainder of this analysis.

\begin{figure}
\includegraphics[width=60mm, angle=90]{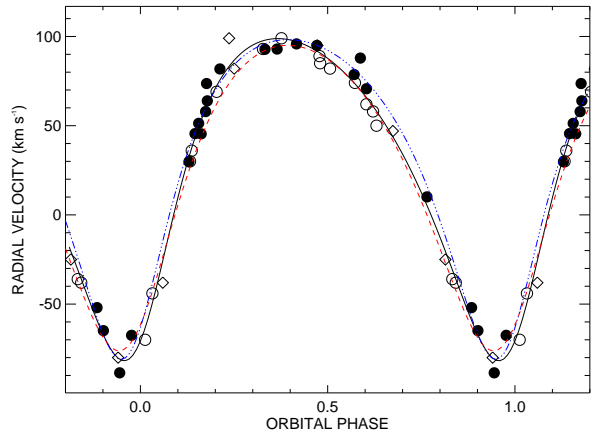}
\caption{\label{sb1} The revised single-lined spectroscopic orbit with data sources marked. Phase 0 indicates the periastron passage.  
New data are shown as solid points, the open circles are from L13, and the open diamonds are from Muratorio et al.~(2008). The black line represents the combined fit of all data, while the blue dash-dotted line is the fit from only our measurements, and the red dashed line is the solution from L13.
 }
\end{figure}

\begin{table*}
\centering
\begin{minipage}{170mm}
\caption{Orbital Elements \label{elements}}
\begin{tabular}{llcccc}
\hline \hline
Element	&	units		&		L13			&	Our Fit to L13			&   New data			&	Combined analysis \\ 
\hline	
$P$		&	days		&			$60.799977\pm0.000014$		&	$60.799977$ (fixed)		&	$60.729\pm0.028$ 		&	$60.753\pm0.003$			\\
$e$		&			&			$0.235\pm 0.003$			&	$0.28\pm0.02$			&	$0.32\pm0.03$			&	$0.29\pm0.01$			\\
$\gamma$ &	km s$^{-1}$ &			$28.4\pm 0.2$				&	$27.7\pm1.0$			&	$35.9 \pm 1.8$			&	$31.2\pm0.8$			\\
$K_1$	&	km s$^{-1}$ &			$85.6$					&	$90.3 \pm 2.0$			&	$89.6\pm2.4$			&	$90.3\pm1.2$			\\
$\omega$ &	$^{\circ}$ 	&			$199\pm 1$				&	$206.4\pm4.0$			&	$195.1\pm 6.3$			&	$208.6\pm 0.8$			\\
$T$	&	HJD		&				$2454951.80\pm 0.56$		&	$2454949.75\pm0.60$	&	$2456467.15\pm1.02$	&	$2455618.88\pm0.13$			\\
$f(M)$	&	$M_{\odot}$ &				$3.63$				&	$4.1\pm 0.29$			&	$3.9\pm0.3$			&	$4.0\pm0.3$			\\
$a_1 \sin i$ &	AU		&				$0.465$				&	$0.484 \pm 0.011$		&	$0.475\pm0.014$		&	$0.479 \pm0.010$			\\
$N_{\rm observations}$	&	&		16	&	16 &	20 & 43 \\
$r.m.s.$	&	km s$^{-1}$ &				\ldots				&	3.70					&	7.20					&	6.98 \\
\hline
\end{tabular}
\end{minipage}
\end{table*}

The orbit is well-behaved with a moderate eccentricity ($e=0.29$). The measured value of $\omega$ is such that the primary is in front of the secondary at phase $0.098$, and the secondary is in front of the primary at phase $0.760$. We find that the mass function, $f(M)$ is somewhat large, with a value of $4.0\pm 0.3 M_{\odot}$. 

\section{The Photometric Variations}

In Figure 2, we show the $V-$band photometry from the AAVSOnet telescope (open circles) and the \most satellite (solid points) phased to the periastron and orbital period found with the spectroscopic orbit. We removed the pulsational signature from the \most data for this plot (see Section 4.2). We performed time-series analysis of the ground-based data and found that the period ($P=60.5 \pm 2.0$) we derive from a Lomb-Scargle periodogram (Scargle 1982) was consistent with that of the spectroscopy, so we adopt the spectroscopic period due to the longer duration of the spectroscopic observations and smaller error in the period determination. The larger scatter in Figure 2 results from the 1--2\% measurement errors inherent to the observations, the pulsational variations still present in the AAVSO data, and from possible long-term variability of the system (e.g.~L13). We note that the light curve resembles the variations presented by L13, who used PHOEBE to model the variations as tidally induced variability with partial eclipses. 

\begin{figure*}
\includegraphics[width=100mm, angle=90]{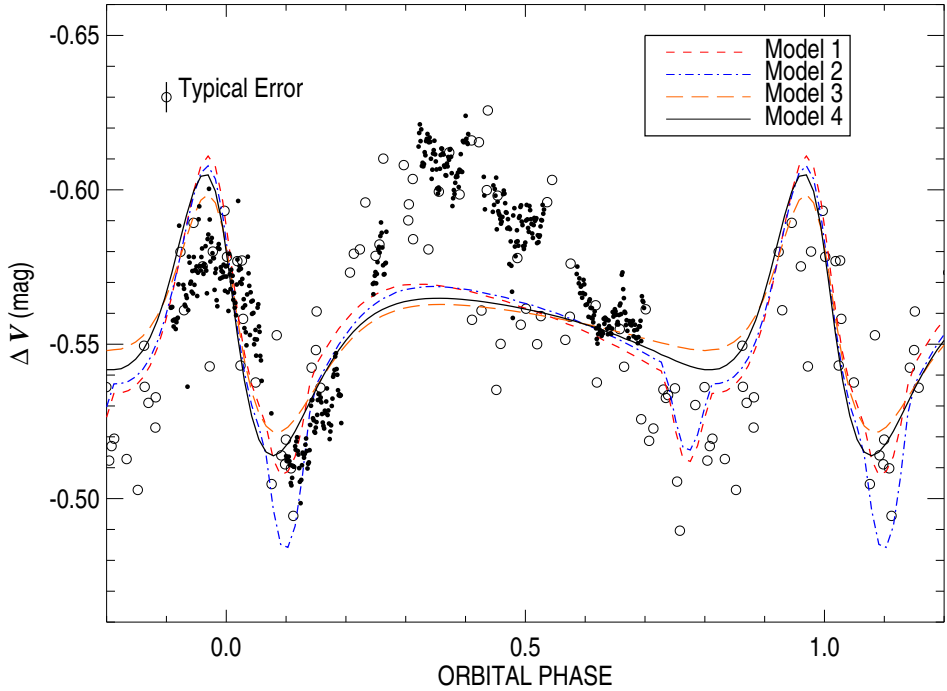}
\caption{\label{phot} $V-$band photometry of MWC 314 as a function of phase. The large open points represent our ground-based data. The solid points are orbital means from \most with the pulsational behaviour subtracted off (see Section 4.2). Our reductions could not remove the instrumental artifacts that were at the end of the observing run at phases 0.7--0.85, so those points are omitted. As the instrumental effects could not be eliminated, we did not use these points in our ellipsoidal fit of the photometry. We overplot our best models of the light curve. The dashed lines represent our best understanding of the ellipsoidal variability for the system (Section 4.1), and correspond to models in Table 2. }
\end{figure*}

\begin{table*}
\centering
\begin{minipage}{70mm}
\caption{PHOEBE Parameters \label{elements}}
\begin{tabular}{lccccc}
\hline \hline
Element				&	units			&	Model \#1	&	Model \#2	&	Model \#3	& Model\#4 \\ 
\hline	
$M_{\rm p}$		&	$M_\odot$	&	39.7		&	39.7		&	61.1	&	4.1	\\
$R_{\rm p}$		&	$R_\odot$		&	81.5		&	81.5		&	94.1	&	35.0	\\
$T_{\rm eff, p}$ 	&	K			&	18,000	&	18,000	&	18,000 & 18,000	\\
$M_{\rm s}$	&	$M_\odot$	&	26.2		&	26.2		&	40.3 &	11.3\\
$R_{\rm s}$	&	$R_\odot$		&	20.6		&	20.6		&	23.7	& 16.4\\
$T_{\rm eff, s}$	&	K			&	6,227	&	12,000	& 	30,000 & 25,000		\\
$i$					&	$^{\circ}$ 		&	73		&	73		&	60	& 60	\\

\hline
\end{tabular}
\end{minipage}
\end{table*}

\subsection{Ellipsoidal Variations}

The photometric variability is likely induced by the star being distorted gravitationally by the companion. This kind of variability has been known and calculated for many years (e.g. Beech 1985) and is referred to as ellipsoidal variability. In recent years, the modeling of eclipsing binaries
has been greatly enhanced by the code PHOEBE (PHysics Of Eclipsing
BinariEs; Pr\v{s}a \& Zwitter 2005). This code gives us the freedom to treat
this system in a variety of different ways, which we explored to examine the photometric variability. The different models we tested are tabulated in Table 2.

We began fitting the system through a recreation of the model of L13. We used our newly derived radial velocity
orbit in Section 3 to set $T$, $P$, $e$, and $\omega$, leaving us with masses,
radii, temperatures, and the orbital inclination as free parameters. The
first model (Model \#1 in Table 2) shows ellipsoidal variability and a partial secondary eclipse. This model, along with our other PHOEBE models, suffers from inadequate flux at phases 0.2--0.5, and is over-estimated from phases 0.7--0.8. However, the variability near periastron is of a correct amplitude and shape to reproduce the light curve at those phases.

We then adjusted the effective temperature of the secondary (Model \#2) to have a hotter secondary star. This demonstrates that the primary is the main source of variation in these models, but the primary eclipse depth is increased drastically. With a small change of the inclination (Model \#3), we then found that the eclipses are no longer seen, which is supported by the light curve shape, even if the models do not fully represent the photometry. Further, in Model \#3, we increased the temperature to that of an O dwarf in order to have a more realistic companion star.

Lastly, we attempted a low mass solution (Model \#4) in which the primary has lost mass via Roche lobe overflow onto the secondary (see Section 6.1). These parameters include a primary with a smaller radius than Models \#1--3 and a secondary mass almost three times larger than the primary. With a moderate inclination of 60$^\circ$, we obtain a model light curve with the same basic shape and amplitude. This shows that the mass ratio derived by L13 can only be considered one possible solution, as the light curve can be reproduced with extraordinary changes in the masses and temperatures, and is not dependent on which star is the more massive component.

For these models, we assumed a semi-detached system where the primary fills its Roche lobe at periastron, as any other configuration provided worse fits of the data. In such a system, we see less flux as we look down the orbital
axis (phases 0.76 and 0.10) and observe the strongest flux
at quadrature phases (0.96 and 0.37) as we view the small and
large profiles, respectively, of the distorted star.
Tides in the system attain a maximum near periastron,
and yield a more extreme maximum and minimum flux then, as observed.
Some problems with the PHOEBE results are likely related to the eccentricity of the system. With eccentric binaries, we have differing values of the separation and Roche potentials as a function of phase. This can lead to changes in the state of the binary from detached to semi-detached over the course of an orbit. The physics involved in such a situation is much more complex and detailed models of such binaries are not yet developed. However, the timing of the primary eclipse is well sampled with \most, and we find no strong evidence of an eclipse event, at least not to the extent of the model presented by L13.

\subsection{Pulsational Behavior Discovered with \most}

Our photometry from \most (also overplotted in Fig.~2) offers the most precise photometric time-series of the system ever obtained. Unfortunately, \most
was never intended to reliably extract astrophysical trends on long time-scales, so the binary-induced signal is not reliable. This is seen to be especially true when we de-trended our data using the light curves of several guide stars in the field, and the \most light curve has a spike at phases $\phi \sim 0.7-0.8$. However, when the long-timescale trends are removed from the \most light curve, we immediately see evidence of pulsational behaviour in the star. An analysis of these de-trended data with Period04 (Lenz \& Breger 2005; Fig.~3) found two significant frequencies, which are listed in Table 3. Each period is represented by a sine wave of the form 
$$A \sin (2 \pi f \times t +\phi),$$
where $A$ is the amplitude of the variation in millimagnitudes, $f$ is the frequency in units of day$^{-1}$, $t$ is the time in observed julian day, and $\phi$ is an offset term in radians that allows for differing the peak time of the sinusoidal wave. These fits explain the data with residuals smaller than one millimagnitude for most data points, which is reasonably consistent with the instrumental performance.

\begin{figure}
\includegraphics[width=90mm, angle=0]{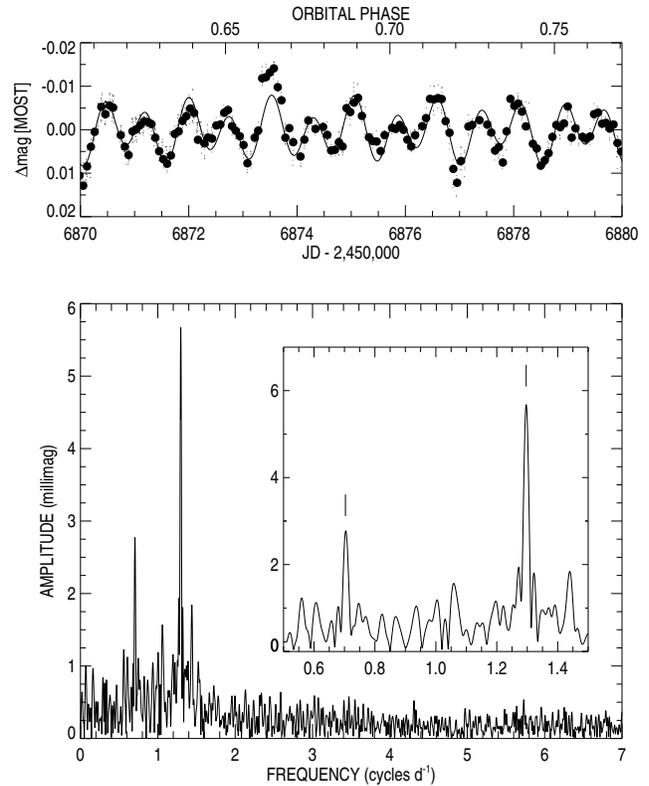}
\caption{\label{MOSTphot} { A portion of the detrended \most photometry of MWC 314 as a function of time and phase is shown in the top panel. Orbital means are shown as large black points, with each individual measurement shown as a small point. Our two-frequency fit is shown as a solid line in this plot. In the bottom panel, we show the Fourier Transform calculated with Period04, with a zoom on the two frequencies found in the inset panel. The full \most light curve is shown in Fig.~A2.}  }
\end{figure}

\begin{table*}
\centering
\caption{Frequencies found with \most. }
\begin{tabular}{lllll}
\hline
Frequency	&	$A$ (mmag)	&	$f$ (cycles d$^{-1}$)	&	$\phi$ (radians)			& $P$ (d)		\\ \hline

1			&	5.59$\pm$0.22	&	1.2964$\pm$0.0004	&	0.3847$\pm$0.0063	&	0.7713$\pm$0.0003	\\
2			&	2.61$\pm$0.22	&	0.7032$\pm$0.0008	&	0.6940$\pm$0.0135	&	1.4221$\pm$0.0016	\\
\hline
\end{tabular}
\end{table*}

These periods are remarkably short for an LBV, where results of other stars have revealed periods of order several days (see, e.g. van Genderen 2001). However, the blue supergiant HD 163899 (B2Ib/II) was studied by Saio et al.~(2006) who found periods of similar duration to MWC 314. L13 have parameters for MWC 314 that indicate a radius of the primary star to be $\sim 80R_\odot$, but the blue supergiant with similar periods (HD 163899) has a radius $\sim 16 R_\odot$. We will further discuss this later in the paper, but the derived radius by L13 is likely much too large to support such short-period pulsations in MWC 314.

\section{Interferometric Results}

The distance of MWC 314 was derived by Miroshnichenko et al.~(1998) to be 3.0$\pm$0.2 kpc using a radial velocity of $+55$ km s$^{-1}$ and the kinematical model of Galactic rotation by Dubath et al.~(1988). We can now adjust this distance by using the $\gamma$ velocity of the single-lined orbit we derived in Section 3 ($\gamma = 31 \pm 1$ km s$^{-1}$). From this derived velocity, we find a closer distance to the system of only 2.4$\pm$0.1 kpc. 
We then derive a luminosity from the expected $V$ magnitude of 9.9, the effective temperature of 18,000 K (e.g. Miroshnichenko et al.~1998, L13), a reddening of $A_V = 5.7$ mag (Miroshnichenko 1996), and a bolometric correction of $-1.7$. The luminosity then becomes $\log (L/L_{\odot}) = 5.7$, with $M_V = -7.8$ and $M_{\rm BOL} = -9.5$, neglecting any effect of the companion and circumstellar and circumbinary material. This result agrees with that of L13, and indicates that MWC 314 is a near twin of the early B-type hypergiant and prototypical LBV, P Cygni, especially with a derived radius of 73$R_\odot$. We caution that if the absorption lines used for the orbit form in the outflow, then the $\gamma$ velocity could be blue-shifted in our line of sight, and this would change this distance estimate.

Our consideration of the CHARA Array results begins with estimates for
the angular size of the visible star and binary orbit.  We estimated
stellar angular diameter by comparing the observed flux distribution $f_\lambda$
with that for a model photosphere $F_\lambda$,
$${{f_\lambda}\over{F_\lambda}} = \left({{R_\star}\over{d}} \right)^2 ~10^{-0.4 A_\lambda}
= {{1}\over{4}} \theta_{LD}^2 ~10^{-0.4 A_\lambda}$$
where $\theta_{LD}$ is the limb darkening angular diameter and $A_\lambda$
is the wavelength dependent extinction.  We made this comparison in
the ultraviolet and optical parts of the spectrum where the observed
flux is probably dominated by the contribution from the visible star.
We followed the example of Miroshnichenko (1996) who used UV spectra from
the {\it International Ultraviolet Explorer} archive and optical magnitudes
to set the observed flux estimates.  We assumed a flux model from the
solar abundance grid of R.\ Kurucz for atmospheric parameters appropriate
for the visible star, $T_{\rm eff} = 18000$~K and $\log g = 2.5$.
We used the extinction law from Fitzpatrick (1999) to set the
extinction law $A_\lambda$ as a function of the reddening $E(B-V)$ and
ratio of total-to-selective extinction $R$. Then a fit of the observed
fluxes with the relation above was made with parameter values of
$E(B-V)=1.81 \pm 0.02$ mag, $R = 3.05 \pm 0.05$, and
$\theta_{LD} = 0.24 \pm 0.02$ mas.  The first two parameters agree
within uncertainties with the results of Miroshnichenko (1996).
In the following analysis we will assume a uniform disc model for
the visible star with an equivalent angular size of
$\theta_{UD} = 0.23$~mas, very slightly smaller than the
limb darkened disc size in the $K$-band (Davis et al.\ 2000).
We argue below (Section 6.1) that the angular size of the orbit is
also quite small ($\approx  0.5$ mas) and that the companion may be
faint because it is hidden in an obscuring gas torus.  Consequently,
in this section we will make the simplifying approximation that the
entire binary system may be represented by a uniform disc of a
size below the resolution limit of our observations.

We collected a large number of squared visibility measurements $V^2$
with varying baselines (45 -- 321 m) that are well sampled on the
$(u, v)$ plane (Fig.\ A1).  Before we attempted to model the visibilities,
we examined observations that were obtained with the CLIMB
beam combiner, which also gave estimates of the closure phase.
These measurements give an indication of the degree
of non-axial asymmetry in the data. All measured triple products have values close to zero, as shown in Figure 4. With the small values of the closure phase, we can assume that the light distribution shows an axial symmetry within our errors at the resolution probed with these interferometric observations. 

 \begin{figure}
\includegraphics[width=60mm, angle=90]{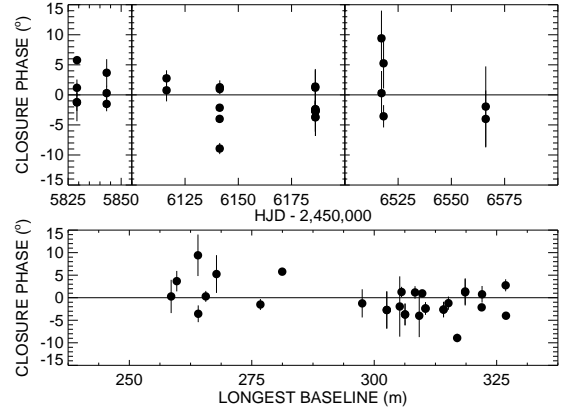}
\caption{\label{fig5} Measured closure phases from the CHARA measurements as a function of time (top panel), where each panel represents a different observing period. The bottom panel represents the same measurements as a function of the longest baseline used. The measurements are all very near zero, so we assume that the system is either spherically symmetric or that the circumbinary disc has axial symmetry at the resolution of the observations.}
\end{figure}

Our adopted interferometric model incorporates the central binary as a uniform disc that is surrounded by a circumbinary disc that is modelled by an elliptical Gaussian on the sky. Any visibility modulation from the binary should be within the errors of the measurements given the small semi-major axis on the sky and the probable faintness of the secondary with respect to the primary. Our model, based upon the methods of Schaefer et al.~(2010) and Touhami et al.~(2013) has six different fitting parameters, and we have 107 measurements of $V^2$. The parameters include 
the size of the central source modeled as a uniform disc (UD),
the flux from the central source (UD Flux),
the flux from the circumbinary disc (1 -- UD Flux),
the full width at half maximum (FWHM) of the major axis,
the FWHM of the minor axis,
and the position angle of the major axis of the circumbinary disc on the sky.
From the measurement of the major and minor axes, we can compute a disc normal inclination angle from the thin disc approximation.
We tabulate the results of various fits in Table 4. We also note that the additional source of error from the correction from the emission line contamination on the interferometry lowered the reduced $\chi^2$ of these models due to the slightly larger measurement errors in the data.

\begin{table*}
\centering
\begin{minipage}{180mm}
\caption{Derived Interferometry Model Parameters. If no error is given, the parameter was held constant. \label{table4}}
\begin{tabular}{llllllll}
\hline
Model		&	UD (mas)	&	UD Flux		& Major FWHM (mas)	&	Minor FWHM (mas)	& 	Position Angle ($^{\circ}$)	& $i$ ($^{\circ}$) & $\chi^2_{\rm red}$	\\
\hline
0	&	$0.08\pm0.88$		&	$0.67\pm0.03$		&	$1.33\pm0.24$	&	$1.33\pm0.24$	&	\ldots		&	\ldots	& 4.24 \\
1	&	$0.49\pm0.14$		&	$0.76\pm0.05$		&	$2.69\pm0.75$	&	$0.45\pm0.51$	&	$141.9\pm4.2$	&	$80.4^{+9.6}_{-20.0}$		& 3.89 \\
2	&	$0.23$			&	$0.70 \pm 0.03$	&	$1.63\pm0.49$	&	$1.23\pm0.22$	&	$114.7\pm28.0$	&	$61.5 \pm 20.0$	& 4.15 \\
3	&	$0.23$			&	$0.69\pm0.02$		&	$1.37\pm0.27$	&	$1.37\pm0.27$	&	\ldots			& 	\ldots	&	4.25 \\
\hline
\end{tabular}
\end{minipage}
\end{table*}

Model \#0 in Table 4 is for a spherically symmetric outflow. The derived uniform disc diameter (UD) for the star is highly unconstrained, {but the resulting $\chi^2$ indicates that the model is on the right track.}
We explored alternate geometries with an elongated wind or circumbinary disc structure. 
Model \#1 allows all parameters to vary and we derive a nearly edge-on disc with a disc normal inclination
that is somewhat larger (but equal within uncertainties) than the
the orbital inclination derived from the light curve in Section 4.1.

The next model (\# 2) fixed the size of the central source to be that of the photospheric size of the primary star as derived from the spectral energy distribution. This is a reasonable choice if we assume little flux emergent from the secondary star. We note that $\chi^2_{\rm red}$ is statistically indistinguishable from the models \#0 and \#1. Lastly, we explored a case where we fixed the central source size, but fit the shell as spherically symmetric (Model \#3).

It is also possible that we are not actually resolving a circumbinary shell or disk, but rather the binary itself. We explored the possibility that we were actually resolving the binary by an examination of the densest observation set obtained on 2012 August 01 (HJD 2,456,141), where we obtained 5 measures of the closure phase, and 15 measures of $V^2$ with CLIMB. We established a grid-based $\chi^2$ minimization where we calculate a $\chi^2$ statistic for a large grid of separations in right ascension, declination, and a flux ratio between the two stars. On this night, our best fit was with a separation of 0.697 mas, a position angle of $263^\circ$, with a flux ratio between the two stars of 0.997. The reduced $\chi^2$ statistic was 22.3, meaning our model did not reproduce well the observations. Similar results were seen for the night of 2012 September 16. These ``fits" actually provide a $\chi^2$ statistic much worse than that of a circumbinary shell or disk. We further note that a flux ratio near unity shows that the resolved companion contributes similar levels of flux as the primary, and that the putative separation is on the same order as the resolving limit of the CHARA Array in the $K'$ band. This flux would likely imply that the orbit would be easily seen to be a double-lined binary, making this result more unlikely.

 \begin{figure}
\includegraphics[width=60mm, angle=90]{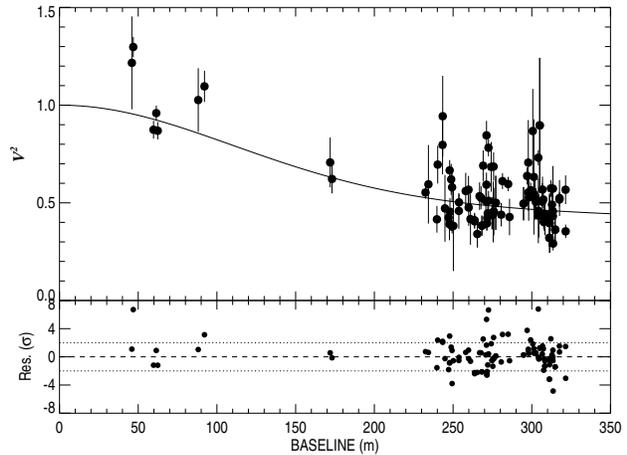}
\caption{\label{fig5} $V^2$ Measurements from the CHARA measurements. We overplot the theoretical visibility curves for the visibility from the spherically symmetric model \#3. Most points fall within 2$\sigma$ of this curve, which is shown in the bottom panel. }
\end{figure}

All of the best $V^2$ models show that we are resolving a circumbinary shell, but it is unclear if the shell is elongated or spherical. In fact, the models we explored cannot distinguish between them, as all the $\chi^2$ values are similar. Further, it may be that our errors in the measurements of $V^2$ are underestimated, making these fits reasonable for these data. 
The spherical model with a fixed UD is shown in Fig.~5, where we compare the measured and calculated visibilities. In this figure, we compare the residuals in the bottom panel by calculating the $(O-C)$ divided by the measurement error, $\sigma_{V^2}$. This shows that the model is reproducing the data within $\sim 2\sigma$ for most data points. As this model seems to adequately fit the $V^2$ measurements, we adopt this for the remainder of the analysis. However, we suspect that better data collected in the future may resolve a multi-component model that includes the central star(s) with the stellar wind and the circumstellar disk seen in spectroscopy.

\section{Discussion}

This study has amassed a large dataset that utilised several observing techniques including spectroscopy, photometry, and interferometry. The most exciting results relate to the fundamental parameters of the system, the precision photometry obtained with {\it MOST} which allowed us to identify pulsational modes in this system, and the exploratory interferometry. 

\subsection{Fundamental Parameters}

Even though the spectroscopic orbit only shows evidence of the primary star, we are able to determine the mass function and mass sum as
$$ M_1 + M_2 = (4.0\pm0.3) M_{\odot} (1 +{{1}\over{q}})^{3} \sin ^{-3} i,$$
where $q=M_2/M_1$ and subscripts 1 and 2 denote the primary (visible) star and secondary star, respectively. 
The mass relation is shown in Figure 6 where we plot $M_2$ as a function of $M_1$ from this equation for three different
values of the inclination ($i=30^\circ$, $60^\circ$, and $90^\circ$).
We showed in Section 4.1 that partial eclipses would appear in
the orbital light curve for $i > 70^circ$ that are not observed.
Furthermore, we assumed the maximal tidal distortion possible by
setting the radius so large that the star fills the Roche surface at
periastron, and consequently any lower inclination would yield
a model with a light curve amplitude that was too small, because
the size of the tidal modulation varies approximately as $\sin i$.
Consequently, we suggest that the orbital inclination probably
falls within the range from $i=50^\circ$ to $i=75^\circ$.

 \begin{figure}
\includegraphics[width=60mm, angle=90]{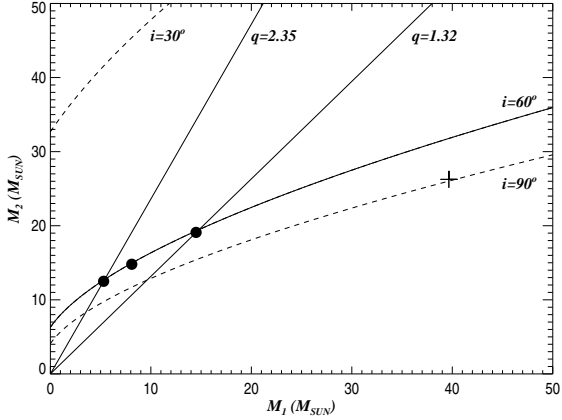}
\caption{\label{fig5} Masses as a function of inclination and mass ratio. The solid black curve represents the mass function at $i=60^{\circ}$. The dashed lines represent $i=30^\circ$ (top) and $i=90^\circ$ (bottom). L13 reported the values shown by a plus sign.
We overplot lines for several assumed mass ratios (from Table 5),
and filled circles show the corresponding masses for $i=60^\circ$.  }
\end{figure}

Sepinsky et al.\ (2007a, 2009) have investigated
how mass transfer eccentric orbit binaries can alter the orbital
elements. They show that momentum transfer caused by RLOF can
have a large influence on the orbit.  In particular,
Sepinsky et al.\ (2007a) show that if the mass ratio reversal
has occurred and the mass transfer rate is high, then
the eccentricity can increase with time.  Their Figure 3 shows that if $M_1/M_2 < 0.76$ ($M_1$ is the mass
donor = the primary in MWC 314), then continued RLOF
will yield a positive time derivative of eccentricity.
Thus, if we accept this result, then the fact that we find
$e=0.29$ in MWC 314 must mean that mass transfer has proceeded
beyond mass ratio reversal and that the current mass ratio obeys this limit, so that $q \geq 1/0.76 = 1.32$. We show this limit in a solid line on Fig.~6, so that any mass solution must fall to the left of this line diagram.

We argued in Section 4.1 that the tidal modulation
of the light curve is best fit if we assume that the visible star
fills its Roche lobe at periastron.  Because the tides are strongest
at periastron, we might also assume that the spin of the star becomes
synchronized with the instantaneous orbital rate at that instant.
We may use these assumptions to explore the consequences of the
Roche geometry for the binary mass ratio.  We can estimate the mass
ratio by comparing the projected rotational velocity, $v \sin i$, to the orbital semi amplitude, $K_1$ in the manner of Gies \& Bolton (1986) with the expression 
$${{v \sin i}\over {K_1}} = \rho \Omega \left(1 + {{1}\over{q}}\right) (1-e^2)^{1/2} \Phi$$
where $q=M_2/M_1$. This expression relates the size of the visible star 
to the Roche radius $\Phi=\Phi (q)$ (Eggleton 1983) 
through a fill-out factor $\rho$ ($=1$ for a Roche filling star), and the angular
rotational rate is expressed relative to the mean orbital synchronous rate through factor $\Omega$. We estimate $v \sin i$ to be $\sim 50$ km s$^{-1}$ based upon the FWHM of the absorption profiles seen in our high resolution spectroscopy obtained with high S/N, such as those obtained with the CTIO 1.5 m and CHIRON, consistent with the value reported by L13. Based upon the analysis of the light curve (Section 4.1)
and the abundant evidence of mass transfer and mass loss in MWC~314,
we assume that the star has a radius that fills its Roche lobe at the periastron passage, so the fill-out factor is given by $$\rho = (1-e)=0.71\pm0.02.$$ If the stars are synchronous as the tidal forces peak at periastron passage, then $$\Omega = {{(1+e)^{1/2}}\over{(1-e)^{3/2}}}.$$
We evaluated the remaining term for the Roche radius $\Phi$ above using
the general expressions for a star in an eccentric orbit from
Sepinsky et al.\ (2007b).  Then we can use the formula above for the
observed ratio $(v \sin i) / K_1$ to find the mass ratio $q=2.36$ for
the case of spin synchronization at periastron ($\Omega = 1.90$).
This mass ratio yields masses of $M_1 \sim 5.3 M_{\odot} $ and $M_2 = 12.5 M_{\odot}$ for intermediate inclination of $i=60^\circ$. Note that if some of the line broadening is due to macroturbulence
instead of rotational Doppler broadening, then the actual $v \sin i$
will be smaller than assumed and the resulting mass ratio larger 
than estimated above. We show the masses derived through this method for both the limiting cases of synchronous rotation at periastron and the eccentricity-growing limit of $q=1.32$ in Table 5, along with an intermediate case. All of these solutions are based upon $i=60^\circ$ and are marked as solid circles in Figure 6.  These slower spinning model solutions
are particularly relevant, because Sepinsky et al.\ (2010) find that
in some circumstances mass transfer episodes at periastron can result
in gas returning back to the donor and decreasing its spin.

\renewcommand{\thefootnote}{\arabic{footnote}}
\begin{table}
\centering
\caption{{Potential Masses of the MWC 314 system$^\dag$, given $i=60^\circ$.}}
\begin{tabular}{cccccccc}
\hline
$\Omega$ & $q$ &  $M_1$  & $M_2$ &   $R_1$ & d(kpc) & a(AU) & a(mas)\\ \hline
1.90 &   2.36  & 5.3 & 12.5 & 36.2 & 1.41 & 0.79  & 0.56 \\
1.60  &  1.82  & 8.1  &14.8 & 43.0 & 1.67 & 0.86  & 0.51 \\
1.26   & 1.32 & 14.5  &19.1 & 54.6 & 2.12 & 0.97 & 0.46 \\
\hline
\end{tabular}
\linebreak
$^\dag$ To change to another inclination, one can rescale by multiplying by $(\sin 60/ \sin i)^3$ and $(\sin 60/\sin i)$ for masses and semi-major axis, respectively
\end{table}

We can then calculate the physical radius of the primary
star $R_1$ by multiplying the fractional Roche radius times the fill-out
factor and times the semimajor axis.  The derived radii are listed in
Table 5 using an assumed inclination of $i=60^\circ$ to find the semimajor
axis.  We determined the angular diameter of the star in Section 5 from
a fit of the spectral energy distribution, and we can use the angular
size to relate the physical radius to the distance,
$R/R_\odot = 25.7 d({\rm kpc})$.  Table~5 column 6 lists the
distance from this relation based upon the physical radius in column 5.
Columns 7 and 8 list the physical and angular semimajor axis of the
binary, respectively, based upon the masses, period, and distance.
The angular semimajor axis is approximately 0.5~mas in all three
spin cases given in Table 5, which suggests that the binary is
probably unresolved in the CHARA Array interferometric observations
(Section 5).  The luminosity of the visible star is large for
the radii given in Table 5, $\log L/L_\odot \approx 5.1 - 5.5$, and
if this is comparable to the initial luminosity of the star before
mass transfer, then the star probably began life with a mass
of $\approx 20 - 30 M_\odot$ (Saio et al.\ 2013).

All the results suggest that the system has gone through a mass reversal process in addition to the mass-loss from stellar winds and eruptions that caused the large ejecta seen by Marston \& McCollum (2008).
Our derived mass ratio would imply that the system is still early in the process of mass transfer, as it has yet to reach the more extreme values found in systems such as HDE 326823 ($M_2/M_1 = 5.3$) or RY Scuti ($M_2/M_1 = 3.9$), but the mass ratio is still clearly reversed. We also note that the observation of very strong Balmer line emission tells us that the primary star still has some hydrogen, supporting the idea that MWC 314 is not as far along in its binary interaction as the pre-(WN $+$ O) binary, HDE 326823. If the low masses we derive are correct, then the Roche radius of the primary must be smaller than estimated by L13, making the primary less luminous and the distance smaller, as indicated in Table 5. 

 \begin{figure*}
\includegraphics[width=85mm, angle=0]{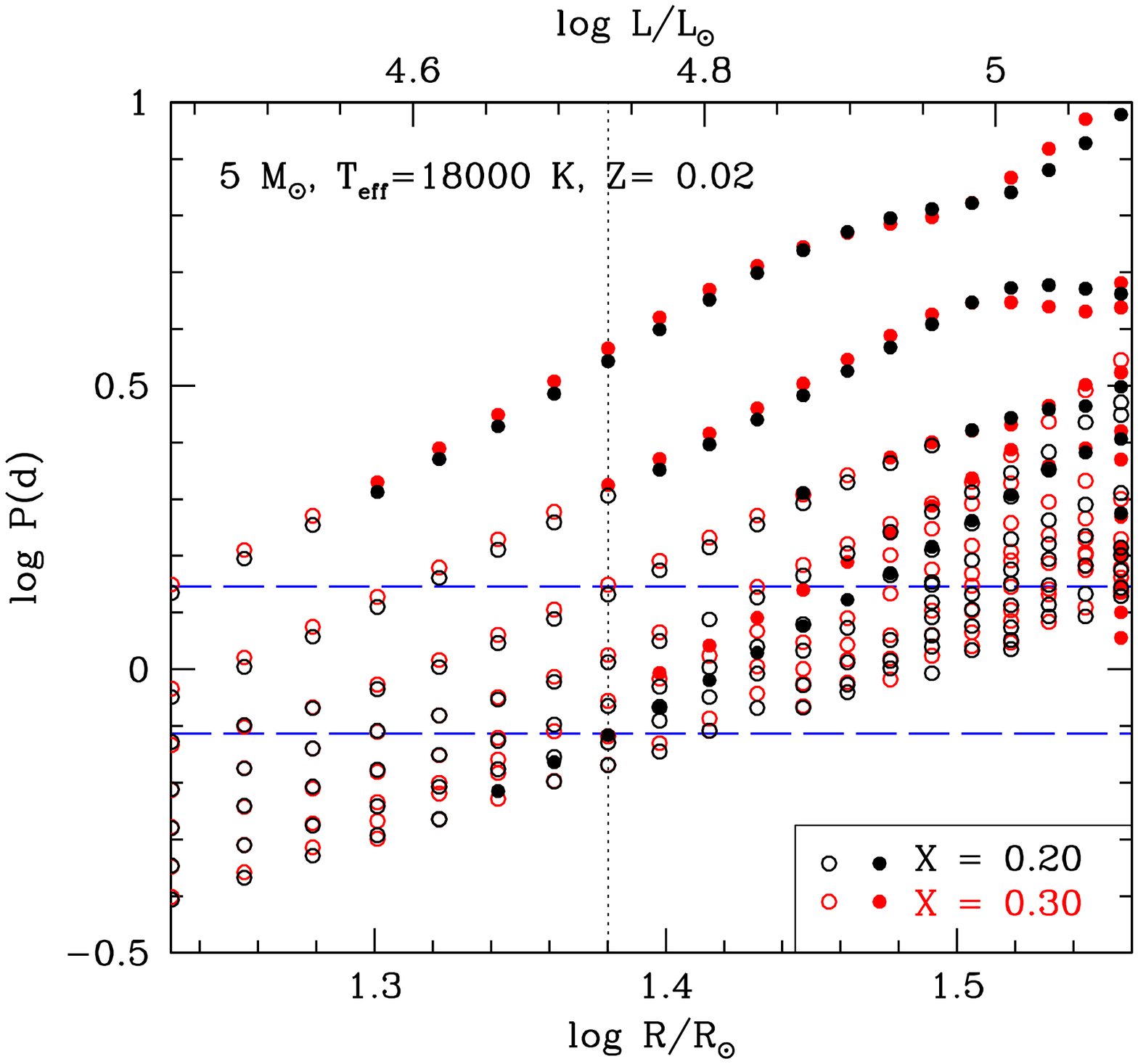}
\includegraphics[width=85mm, angle=0]{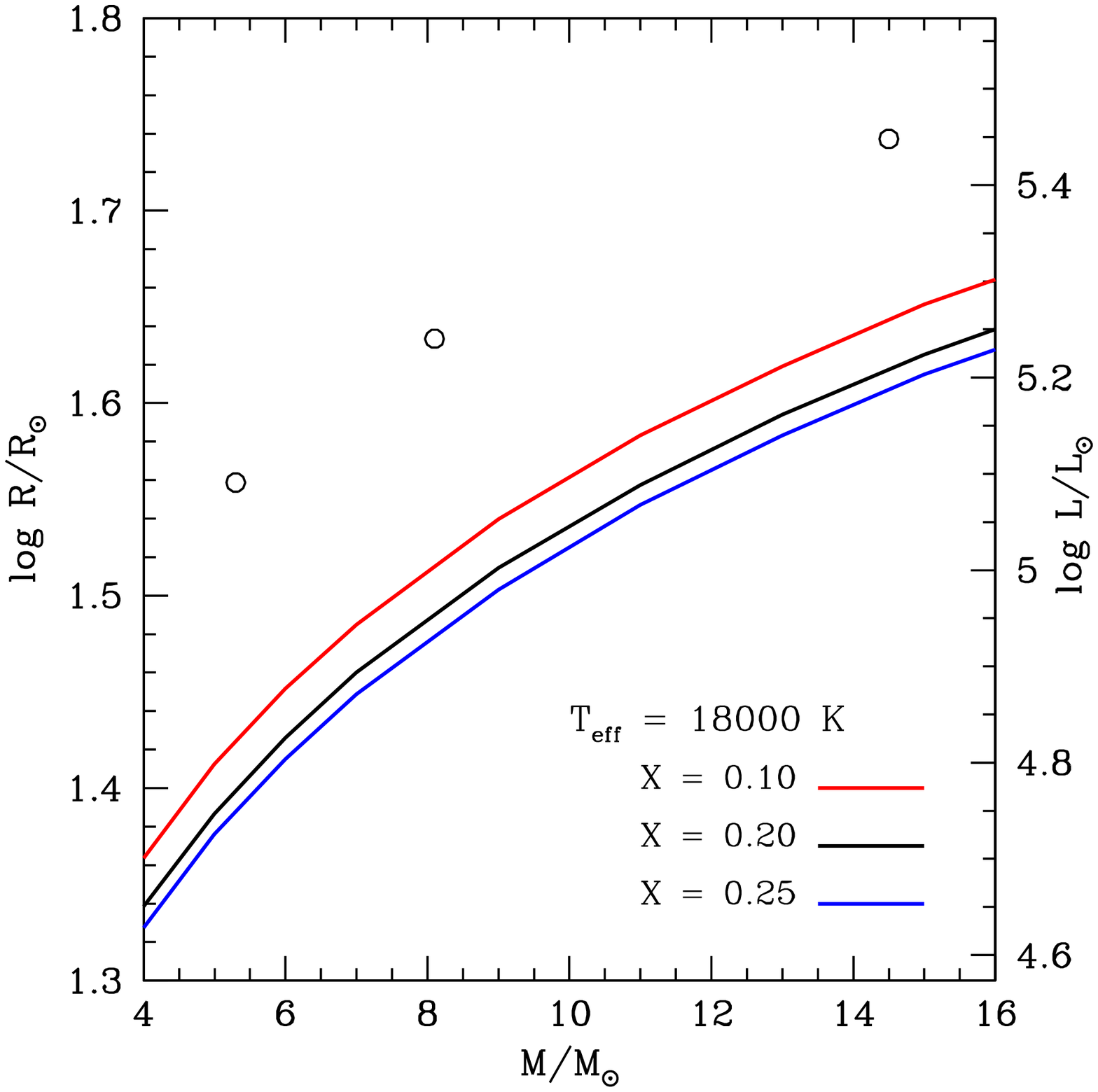}
\caption{\label{figpuls}  A comparison of the observed two periods (0.77 d and 1.4 d; horizontal dashed lines) with theoretical pulsation periods for models of a stellar mass of $5 M_\odot$ and an effective temperature of 18000 K with hydrogen mass fractions X=0.2 (black symbols) and X=0.3 (red symbols). Filled and open circles indicate excited and damped pulsation modes, respectively.  The bottom and the top horizontal axes measure stellar radius and luminosity, respectively. The longest period of each model is the fundamental mode. We find that the observed dominant period (P=0.77d) is reproduced by an excited mode in a $5M_\odot$ model at a radius of 24 $R_\odot$ ($\log R/R_\odot = 1.38$; vertical dotted line) with X = 0.2. The second overtone of this model has a period of 1.4 d close to the observed second period, although the mode is not excited. We note that the model predicts the fundamental mode with a period of 3.5 d to be excited, although no such periodicity is detected from our \most light curves.  On the right panel, we show how the radius and luminosity scale with mass for the range of allowed masses (Section 6.1) and hydrogen fractions that can support this pulsational mode. The open circles denote the values derived by the Roche geometry analysis in Section 6.1 and Table 7.} 
\end{figure*}

The visible star mass that we derive is surprisingly low for a star that
has a spectrum resembling the prototypical LBV, P Cygni.
The masses derived by L13 are much
closer to expectations for an LBV mass, but we caution that
their results are based upon the mass function from the radial
velocity curve and an inclination and mass ratio estimated
from fits of the light curve.  Recall from Section 4.2 and Figure 2
that the light curve is complex and that a PHOEBE calculation with a large
mass ratio $q$ produces a light curve (Model \#4) that is qualitatively
similar to those for a small mass ratio (Models \#1,2,3).
Consequently, there is probably a large range in adopted mass ratios
and hence masses (Fig.~6) that will yield model light curves
similar to the observed photometric light curve.
Our determination of the mass ratio (above) was made
assuming synchronous rotation and a visible star that fills its
Roche lobe at periastron.  L13\ recognised that
their model implied slower than synchronous rotation
for the visible star (they suggested a possible synchronous
relation with gas forming the inner boundary of the circumbinary disc).
However, stars that fill their Roche lobes experience
strong tidal forces that drive the system towards synchronous
rotation on a relatively short time scale, so the masses that
we derive based upon the assumption of synchronous rotation
are worthy of detailed consideration.

Note that we have made the assumption that the absorption
lines we measured form in the atmosphere of the visible star.
We suspect that the mass gainer in the system is surrounded
by a thick gas torus (see below), and it is possible that
the absorption lines form in this torus instead.  If so,
then the radial velocity curve would apply to the mass gainer
and the identities of the stars would be swapped in the mass-mass
diagram (Fig.~6).  The measured ``projected rotational velocity''
in this case would correspond to Keplerian motion in the gas
disc surrounding the mass gainer and not to the rotational
broadening of the mass donor star, so the $(v \sin i)/K_1$
argument would not apply.   However, spectral lines formed
in a torus surrounding the mass gainer have been detected for both
$\beta$~Lyr (Ak et al.\ 2007) and RY~Scuti (Grundstrom et al.\ 2007),
and in both these cases the lines display very large rotational
broadening.  Consequently, we suspect that the narrow absorption
lines we observe in the spectrum of MWC~314 do not form in
gas torus but are associated with the photosphere of the visible star.

\renewcommand{\thefootnote}{\arabic{footnote}}
\begin{table*}
\centering
\caption{ Models having an excited 0.77 d pulsation mode ($X=0.2$, $Z=0.02$, $T_{\rm eff} = 18,000$ K). }
\begin{tabular}{ccccc}
\hline
Mass ($M_\odot$)		& 	Radius ($R_\odot$)	& $\log (L/L_\odot)$	&	Other excited period(s)	&	Period closest to 1.40 d$^\dag$\\ \hline
4.0	&	21.5	&	4.64	&	1.98, 3.47 d	&	1.32 d			\\
5.0	&	24.0	&	4.73	&	3.50 d		&	1.36 d \\
6.0	&	26.3 	&	4.81	&	3.52 d		&	1.39 d \\ 
7.0	&	28.5	&	4.88	&	3.58 d		&	1.43 d \\
9.0	&	32.3	&	4.99	&	3.59 d		&	1.48 d \\
\hline
\end{tabular}
\linebreak
$^\dag$ Note that these periods are all second overtones, and all are damped.
\end{table*}

\subsection{Pulsational Behaviour}

 \begin{figure}
\includegraphics[width=60mm, angle=90]{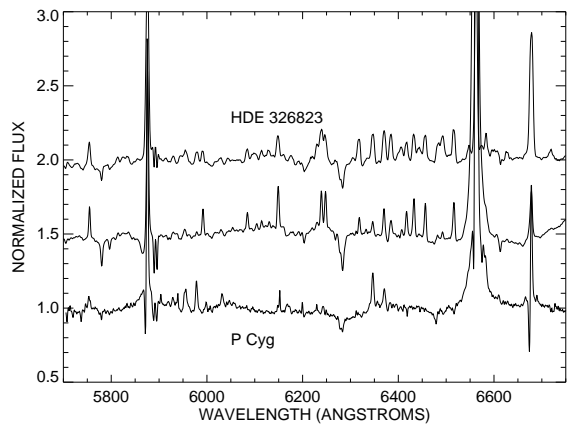}
\caption{\label{fig5} A comparison between the spectrum of MWC 314 (bold, center), HDE 326823 (Richardson et al.~2011; offset for clarity), and P Cygni (Richardson et al.~2013, bottom) shows that MWC 314 and HDE 326823 are nearly spectroscopic twins with the exception of the stronger hydrogen emission and lack of a few weak emission lines from MWC 314. Similarly, the comparison with P Cygni shows the similarity of the stellar winds of the two stars.     }
\end{figure}
The {\it MOST} photometry presented us with an opportunity to explore
the pulsational properties of this unusual system.  In particular, a
pulsational analysis may provide another clue that supports the
relatively low mass of the visible star that we estimated
above.  We examined the pulsational stability of MWC~314 using the nonrotating
stellar models and the modal analysis described by Saio et al.\ (2013).
Saio et al.\ discuss how pulsations may be excited among blue supergiants
in cases where the luminosity to mass ratio is large (for example,
after extensive mass loss during a prior red supergiant phase).
This is especially pertinent for MWC~314 if the stellar mass has
decreased significantly through mass transfer while maintaining the
luminosity of the He-burning core.

We began by exploring what range of mass and radius would yield models
with an excited mode pulsational period of 0.77~d for a stellar effective
temperature of $T_{\rm eff }=18000$~K.  We found that there were no
models with normal hydrogen abundances that could support the dominant
pulsational model.  However, if the hydrogen mass fraction $X$ was set to
a value less than 0.3, then we were able to find models with a pulsational
frequency that matched the observed one.  The parameters for these
solution families are illustrated in Figure 7 (right panel) that plots
the stellar mass, radius, and luminosity for several trial values of $X$.
Figure 7 also shows the stellar parameters estimated above from the Roche
geometry arguments (Table 5), and we see that these indicate an
overluminosity for mass relative to the smallest hydrogen fraction
family of pulsation models ($X=0.10$).  It is possible that even lower
hydrogen abundance models may excite pulsations like those observed,
but it is difficult to specify the structure of the stellar envelope
for such a stripped down star. This pulsational frequency is unlike frequencies often observed in LBVs, which are both longer time-scales and not strictly periodic (e.g., van Genderen~2001). The fundamental 0.77 d period is also shorter in duration than the dominant periods, and smaller in amplitude than the pulsational frequencies reported for 24 B supergiants reported by Lefever, Puls, \& Aerts (2007).

The pulsation mode identified with the 0.77 d period of MWC~314 is a kind
of strange mode associated with the He~II ionization zone
($\log T \approx 4.5$), and the amplitude is  strongly confined to the
outermost layers with $\log T < 4.6$.  On the other hand, longer-period
fundamental and first overtone modes with periods longer than a few days
are excited by the $\kappa$-mechanism around the Fe-opacity peak at
$\log T \approx 5.3$ as occurs in other B-type variables with normal
surface H-abundance.  With a low hydrogen abundance of $X \lesssim 0.25$,
the mode having a period of 0.77 d at $T_{\rm eff} = 18000$~K is excited
by the $\kappa$-mechanism in the He~II ionization zone for models
with masses consistent with those from the orbital analysis, although the
radius required for the period tends to be smaller than that of the
critical Roche Lobe as shown in Figure 7 (right panel).

In Table 6, we summarize the models with $X=0.2$ in which the 0.77 d period
is a supported pulsational mode.  These models show that there is a family
of lower-mass models in which this pulsational frequency is excited.
The left hand panel of Figure 7 shows an example how the periods and
the stability of the pulsation change for models with $M = 5 M_\odot$ as a
function of radius at $T_{\rm eff} = 18000$~K. The model reproduces the
0.77 d period at $\log R/R_\odot = 1.38$. In addition, the period of the
second overtone of these models falls in the range of 1.3 to 1.4 d, which
is similar to the observed secondary period (1.4 d), although the mode is
predicted to be damped.  We note that these models predict that the
fundamental mode with a period of $\approx 3.5$~d should be excited,
but we find no evidence of this periodicity in the {\it MOST} data (Fig.~3).

This pulsational analysis demonstrates that the kind of pulsational
frequencies observed are broadly consistent with models in which the
stellar mass is relatively small and the hydrogen abundance is low.
This provides additional evidence that the visible star is the
stripped-down remnant of binary mass transfer and that it is
currently the lower mass component in the binary (Table 5).

\subsection{Interferometry}

The only LBV with comparable interferometric data is P Cygni (Richardson et al.~2013). 
The best models to re-produce the spherically symmetric wind of P Cyg were created either by the non-LTE code CMFGEN or a simple uniform disc surrounded by a Gaussian halo, and show that the $H-$band flux emerges from a halo about 2.4 times larger than the photosphere. 
In comparison, the emission from the $K-$band wind of MWC 314 is about 6 times larger than the photosphere for the case of MWC 314 (Model \#3 in Table 4).
Despite the differing wavelength of the observations for MWC 314 and P Cygni, the differences between the $H-$ and $K-$ bands are fairly small (a few percent in the case of the CMFGEN model of P Cygni computed by Najarro 2001). We note
that optical polarimetric observations of MWC~314 show some evidence of a
preferred direction (Wisniewski et al.\ 2006), which may in fact be emerging from the circumbinary disc component of the flux, rather than a wind asymmetry.

The largest remaining question is why the sizes are so different between P Cygni and MWC 314. The optical and near-infrared spectra appear similar (L13, Fig.~7), so we may suspect them to appear similar in physical size and geometry. 
However, in the case of an interacting binary many of the existing models for hot star winds may not be good approximations. We know that MWC 314 has a circumbinary disc, and it must account for some of the flux we see with the CHARA Array. However, we had a difficult time discerning different models of the visibilities, other than the extended halo of light surrounding the system. It's large size compared to P Cygni likely indicates that the CHARA Array is seeing evidence of both the wind and the disc. 

The large halo observed with CHARA likely has an
origin in both a large circumbinary disc and a circumbinary wind, and there is evidence from
spectroscopy that the binary is embedded in a disc.
The appearance of strong emission lines throughout the optical
and near-infrared spectrum of MWC~314 indicates
the presence of circumstellar gas from ongoing mass loss.
With the exception of the strong Balmer lines and He I
emission lines, the optical emission lines tend to be weakly
ionised metal lines, such as Fe II. These lines exhibit double-peaked
profiles (Miroshnichenko et al.\ 1998) that are stationary
in radial velocity (L13), indicative of an origin in a circumbinary disc rather than in either star. 
The same double-peaked and stationary emission lines
are found in the spectrum of the spectroscopic binary 
HDE 326823.  Richardson et al. (2011) argue that these 
lines form in a circumbinary disk that is fed by mass loss
from the binary.   The visible star in HDE 326823 
is losing mass to a hidden secondary star via Roche lobe Overflow 
(RLOF) and losing mass to the circumbinary disk by outflow 
through L2.   In fact, the spectral similarities
of the two stars are remarkable. In Figure 7, we
show a comparison of the average low resolution spectrum
of HDE 326823 (Richardson et al. 2011) and a similar resolution
spectrum of MWC 314 we obtained at the Observatoire
du Mont M\'{e}gantic. With the exception of the stronger H$\alpha$
line in MWC 314 and the absence of a few weak emission
lines, the stars can be considered spectroscopic twins. HDE 326823 is an example of a W Serpentis binary (Tarasov 2000), where a less-massive primary star has lost mass onto a secondary star hidden behind an optically thick accretion torus (Nazarenko \& Glazunova 2006). In these systems, mass exchange and systemic mass loss have drastically altered the stellar masses.

\section{Summary and Future Work}

With the recent discovery that the LBV candidate HDE 326823 is an interacting binary and with the work presented here and in L13 on MWC 314, we now have two candidate LBVs that are probably interacting binaries. Plavec (1980) and Tarasov (2000) show that the W Serpentis binaries can have a mass-loss and transfer rate up to $10^{-4} M_{\odot} {\rm yr}^{-1}$, which is comparable to the mass-loss rates of LBVs. However, we caution that mass-loss via a stellar wind and mass transfer are very different processes, which may have very different mass-loss rates. 
The spectral appearance of double-peaked emission lines in these highly
luminous stars may be an observational way to find more interacting
binaries in the future, so that high spectral resolution
time-series observations will help to distinguish
between LBVs and mass-transferring binaries.
MWC 314 and HDE 326823 are the only two
known LBV candidates that show double-peaked emission,
but others may be found in the future as
massive stars identified through infrared surveys (e.g., Wachter
et al.\ 2010; Stringfellow et al.\ 2012) are examined at higher
spectral resolution.  For example,
the supergiant B[e] star (sgB[e]) Wd1-9 may show similar
properties to HDE 326823 and MWC 314 (Clark et al.\ 2013).

In summary, we found the following properties related to the interacting binary MWC 314. 
\begin{enumerate}
\item MWC 314 is a single-lined spectroscopic binary with a period of 60.753 d and a moderate eccentricity ($e=0.29$). The full orbital parameters are given in Table 1.
\item The system shows photometric variability modulated with the orbital period.
We constructed light curve models for the tidal deformation of
the star around periastron that were made with the period,
epoch, eccentricity, and longitude of periastron set from the
spectroscopic results.  The model light curves capture the
main features of the orbital light curve (timing and number of
maxima), but the model over- and under-predicts the flux
just before and after visible star inferior conjunction,
respectively.  We suspect that these differences are related
to wind and/or disc asymmetries that are not included in the
PHOEBE model of the light curve. Solutions similar to that
of L13 can be found with both large and small masses.
\item With the \most photometry, we discovered two pulsational periods in the system. These periods cannot be supported in a star with parameters (mass, radius) of a typical LBV, but can easily represent a hydrogen-poor, low-mass star.
\item From the CHARA Array measurements of the squared visibility of the system, we found that a halo of light around
the binary is partially resolved.  We argue that the angular
size of the halo is too large for the wind alone, and it
probably represents the flux of the wind and circumbinary disc.
\item We demonstrated how a consideration of the
Roche geometry can be used to derive the mass ratio from the
observed ratio of projected rotational velocity to orbital
semiamplitude (independent of the system inclination).
If we assume that the visible star spins with the same angular
rate as the orbital advance at periastron, then we derive
a mass ratio $q=M_2/M_1=2.36$, indicating that the donor star
is now the lower mass component in the binary.  We consider
other cases in which the spin rate is lower, but the mass
ratio is limited to $q>1.32$ if we accept models that predict
a growth in eccentricity with mass transfer (Sepinsky et al.\
2007a, 2009).  There is a factor of 3 -- 10 discrepancy between the masses of the L13~study and ours. This discrepancy could be resolved with the appropriate observations in the future that spectroscopically determine the secondary radial velocity curve. However, that may not be possible if the secondary is hidden in an accretion torus.
\end{enumerate}

The system presents opportunities to study mass transfer with multiple observing strategies. 
Further efforts should be employed to obtain very high signal-to-noise spectroscopy with high spectral resolution. Such spectroscopy may reveal the nature of the newly-discovered pulsational modes over short time-scales or find a spectroscopic signature of the companion over orbital time-scales.
The analysis of the interferometry may be able to be improved if a near-infrared light curve is measured for the system, which we have begun trying to do with the CPAPIR instrument (Artigau et al.~2004). This would allow us to account for any 
changes with orbital phase in the ratio of central binary 
to surrounding flux, which would lead to an improved 
interpretation of the interferometric results.  A combination of this with the distance to MWC 314 with the recently launched GAIA satellite will provide constraints on the physical size of the outflow, allowing us to better understand mass exchange in this binary system. All of these analyses will yield insights into the physics of the system which likely includes RLOF and accretion, and which could include either a circumbinary disc or jets. MWC 314 is an exciting target for our understanding of post-main sequence binary interactions of massive stars. 

\section*{Acknowledgements}
We thank the anonymous referee for helping the analysis and presentation of this paper. We are grateful to Fred Walter (Stony Brook University) for his scheduling of spectroscopic observations with the CTIO 1.5 m, to the CTIO SMARTS staff for queue observing support, and to Todd Henry (Georgia State University) for assistance in scheduling the initial observations with the SMARTS echelle spectrograph. We are also grateful to John Monnier (Univ.~of Michigan) for contributions to data reduction and analysis. 
We thank Pierre-Luc L{\'e}vesque, Bernard Malenfant, Ghislain Turcotte, and Philippe Vall\'{e}e for their assistance in obtaining data at the Observatoire du Mont M\'{e}gantic. Some spectra with the CTIO 1.5 m were obtained through the NOAO Programs 2009B-0153 and 2012A-0216. 
This work was partially based on observations obtained at the 2.1-m Otto Struve and 2.7-m Harlan. J. Smith telescopes of the McDonald Observatory of the University of Texas at Austin.
This work was also based partially on observations obtained at the Mercator telescopes and HERMES spectrograph of the Instituto de Astrof\'{i}sica de Canarias.
This research was made possible through the use of the AAVSO Photometric All-Sky Survey (APASS), funded by the Robert Martin Ayers Sciences Fund. 
Operational funding for the CHARA Array is provided by the GSU College of Arts and Sciences, by the National Science Foundation through grants AST-0606958, AST-0908253, and AST-1211129, by the W. M. Keck Foundation, and by the NASA Exoplanet Science Institute. We thank the Mount Wilson Institute for providing infrastructure support at Mount Wilson Observatory. The CHARA Array, operated by Georgia State University, was built with funding provided by the National Science Foundation, Georgia State University, the W. M. Keck Foundation, and the David and Lucile Packard Foundation.
This research has made use of the SIMBAD database, operated at CDS, Strasbourg, France.

NDR is grateful for his CRAQ (Centre de Recherche en Astrophysique du Qu\'ebec) postdoctoral fellowship. 
AFJM and NSL are grateful for financial support from NSERC (Canada) and FRQNT (Qu\'{e}bec).
DRG and GS acknowledge support from NSF grant AST-1411654.
AM and SZ acknowledge support from DGAPA/PAPIIT project IN100614.
TSB acknowledges support provided through NASA grant ADAP12-0172.

\appendix

\section{Supplementary Material}

\begin{table*}
\centering
\begin{minipage}{180mm}
\caption{Spectroscopic Observing Log \label{table1}}
\begin{tabular}{lllccll}
\hline
Telescope		&	Spectrograph						&  Range (\AA)	&	$R$		& 	$N_{\rm spectra}$	&	HJD (first)		&	HJD (last)	\\
\hline
Struve 2.1 m	&	Sandiford Cassegrain Echelle			&	5600--6800			&	60,000	&	2				&	2452192.6	&	2452542.6 \\
Harlan Smith 2.7 m & TS2 Echelle Spectrograph 			& 3600--9875				 &	60,000	&	1				&	2455082.7	& \ldots \\
CTIO 1.5 m	&	SMARTS Fiber echelle				&	4800-7200			&	40,000	&	7				&	2455429.6	&	2455449.5 \\

CTIO 1.5 m	&	CHIRON echelle					&	4500--8500			&	28,000	&	4				&	2456029.9	&	2456079.8 \\
Mercator 1.2 m	&	HERMES Spectrograph				&	3800--8750			&	50,000	&	3				&	2456053.6	& 	2456133.5 \\
OMM 1.6 m	&	Cassegrain, 1200 g mm$^{-1}$ grating 	&	4500--6700			&	3,200	&	2				&	2456489.6	& 	2456494.7 \\
San Pedro Martir 2.1 m &	Echelle REOSC				&	4600--8100			& 	18,000	&	1				&	2456587.6	& \ldots \\
\hline
\end{tabular}
\end{minipage}
\end{table*}

\begin{table*}
\centering
\begin{minipage}{180mm}
\caption{CHARA Interferometric Observing Log \label{table2}}
\begin{tabular}{llllll}
\hline
UT Date		&	Beam Combiner	&	Baseline(s)		& Baseline Length(s)	 [m]	&	N$_{\rm observations}$	& 	Calibrator HD number(s)		\\
\hline
2010 Sep 04	&	Classic			&	S2/E2			&	248				&	1					&	182101					\\
2010 Sep 05	&	Classic			& 	S2/W2, W1/W2		&	177,107			&	2, 2					& 	174897	\\
2010 Sep 21	&	Classic			& 	E1/E2			&	65				&	2					& 	182101	\\
2010 Sep 22	&	Classic			& 	E1/E2			&	65				&	2					& 	174897	\\
2011 Sep 25	&	CLIMB			& 	S2/W1/E1			&	249, 313, 302		&	4					& 	182101	\\
2012 Jul 06	&	Classic			& 	S1/E1			&	330				&	6					& 	182101, 184606	\\
2012 Jul 08	&	CLIMB			& 	S1/E1/W1			&	330, 313, 278		&	2					& 	182101, 184606	\\
2012 Aug 02	&	CLIMB			& 	S1/E1/W1			&	330, 313, 278		&	5					& 	182101	\\
2012 Sep 16	&	CLIMB			& 	S1/E1/W1			&	330, 313, 278		&	5					& 	182101, 184606	\\
2012 Sep 22	&	CLIMB			& 	S1/E1/W1			&	330, 313, 278		&	5					& 	182101, 184606	\\
2013 Aug 13	&	CLIMB			& 	S1/E2/W1			&	278, 251, 278		&	2					& 	174897, 182101	\\
2013 Aug 13	&	Classic			& 	S1/E2			&	278				&	1					& 	174897	\\
2013 Oct 01	&	CLIMB			& 	S1/E1/W1			&	330, 313, 278		&	2					& 	182101, 184606	\\
\hline
\end{tabular}
\end{minipage}
\end{table*}

\begin{table*}
\centering
\begin{minipage}{160mm}
\caption{New Radial Velocity Measurements \label{table-rv}}
\begin{tabular}{llll}

HJD	-2,450,000	&	$V_r$ (km s$^{-1}$)		&	Phase &	Telescope \\ 
\hline

       2192.6045 &        $+$70.7 &       0.604 & McDonald Observatory \\
       2542.6229 &        $+$93.0 &       0.365 & McDonald Observatory \\
       5082.6760	&	 $+$57.9	&	0.174 & McDonald Observatory \\
       5429.5890 &        $-$52.0 &       0.884 & CTIO SMARTS echelle \\
       5430.6169 &        $-$64.9 &       0.901 & CTIO SMARTS echelle \\
       5444.4983 &        $+$29.8 &       0.130 & CTIO SMARTS echelle \\
       5445.4805 &        $+$45.6 &       0.146 & CTIO SMARTS echelle \\
       5446.4785 &        $+$45.4 &       0.162 & CTIO SMARTS echelle \\
       5447.4750 &        $+$63.9 &       0.179 & CTIO SMARTS echelle \\
       5449.5198 &        $+$81.8 &       0.212 & CTIO SMARTS echelle \\
       6029.8825 &        $+$10.1 &       0.765 & CTIO CHIRON \\
       6053.6094 &        $+$51.3 &       0.156 & Mercator-HERMES \\
       6054.8914 &        $+$73.6 &       0.177 & CTIO CHIRON \\
       6078.8322 &        $+$78.7 &       0.571 & CTIO CHIRON \\
       6079.8378 &        $+$87.9 &       0.587 & CTIO CHIRON \\
       6103.4839 &        $-$67.5 	&       0.977 & Mercator-HERMES \\
       6133.5428 &        $+$95.0 &       0.471 & Mercator-HERMES \\
       6489.6439 &        $+$92.8 &       0.333 & Observatoire du Mont M\'egantic \\
       6494.7573 &        $+$95.8 &       0.417 & Observatoire du Mont M\'egantic \\
       6587.5821 &        $-$88.5 &       0.945 & SPM echelle \\
\hline
\end{tabular}
\end{minipage}
\end{table*}

\begin{table*}
\centering
\begin{minipage}{160mm}
\caption{Ground-based Photometric Measurements \label{table-phot}}
\begin{tabular}{lllllllll}

HJD	-2,450,000	&	$B$		&	$\sigma_B$ &	$V$		&$\sigma_V$ 	&	$R$		& $\sigma_R$	&	$I$		& $\sigma_I$ \\ 
\hline
5275.011	&	11.899	&	0.008	&	10.135	&	0.005	&	8.607	&	0.004	&	7.437	&	0.004 \\
5280.941	&	11.892	&	0.007	&	10.116	&	0.016	&	8.578	&	0.004	&	7.441	&	0.008 \\
5291.996	&	11.904	&	0.006	&	10.153	&	0.004	&	8.615	&	0.004	&	7.486	&	0.004 \\
5294.886	&	11.964	&	0.007	&	10.173	&	0.010	&	8.624	&	0.032	&	7.491	&	0.009 \\
5298.875	&	11.960	&	0.012	&	10.181	&	0.029	&	8.630	&	0.012	&	7.492	&	0.024 \\
5312.838	&	11.916	&	0.006	&	10.141	&	0.012	&	8.598	&	0.009	&	7.465	&	0.005 \\
5321.808	&	11.972	&	0.019	&	10.206	&	0.059	&	8.650	&	0.013	&	7.491	&	0.011 \\
5324.800	&	11.955	&	0.007	&	10.180	&	0.005	&	8.599	&	0.014	&	7.471	&	0.010 \\
5327.801	&	11.897	&	0.011	&	10.143	&	0.048	&	8.609	&	0.014	&	7.444	&	0.007 \\
5330.792	&	11.878	&	0.024	&	10.134	&	0.025	&	8.588	&	0.036	&	7.449	&	0.018 \\
5333.778	&	11.897	&	0.006	&	10.121	&	0.037	&	8.581	&	0.031	&	7.429	&	0.004 \\
5337.879	&	11.889	&	0.007	&	10.104	&	0.004	&	8.568	&	0.012	&	7.422	&	0.004 \\
5340.898	&	11.890	&	0.005	&	10.101	&	0.018	&	8.569	&	0.005	&	7.411	&	0.010 \\
5352.873	&	11.962	&	0.016	&	10.178	&	0.009	&	8.641	&	0.007	&	7.495	&	0.004 \\
5359.879	&	11.973	&	0.005	&	10.183	&	0.011	&	8.648	&	0.012	&	7.493	&	0.004 \\
5362.871	&	11.979	&	0.005	&	10.186	&	0.022	&	8.650	&	0.010	&	7.505	&	0.014 \\
5365.869	&	11.966	&	0.017	&	10.203	&	0.004	&	8.647	&	0.024	&	7.509	&	0.010 \\
5368.826	&	11.975	&	0.005	&	10.183	&	0.025	&	8.656	&	0.004	&	7.505	&	0.014 \\
5382.796	&	12.018	&	0.015	&	10.222	&	0.015	&	8.666	&	0.027	&	7.520	&	0.016 \\
5455.716	&	11.898	&	0.019	&	10.132	&	0.009	&	8.607	&	0.004	&	7.455	&	0.019 \\
5468.691	&	11.922	&	0.029	&	10.157	&	0.008	&	8.610	&	0.021	&	7.484	&	0.010 \\
5471.712	&	11.897	&	0.012	&	10.140	&	0.004	&	8.587	&	0.022	&	7.472	&	0.004 \\
5476.705	&	11.942	&	0.007	&	10.161	&	0.008	&	8.633	&	0.019	&	7.490	&	0.015 \\
5479.685	&	11.967	&	0.031	&	10.197	&	0.017	&	8.644	&	0.004	&	7.513	&	0.009 \\
5498.622	&	11.905	&	0.005	&	10.139	&	0.005	&	8.601	&	0.004	&	7.457	&	0.008 \\
5506.596	&	11.943	&	0.020	&	10.168	&	0.010	&	8.626	&	0.004	&	7.489	&	0.004 \\
5511.621	&	11.915	&	0.005	&	10.120	&	0.032	&	8.587	&	0.022	&	7.451	&	0.010 \\
5604.032	&	11.971	&	0.006	&	10.211	&	0.031	&	8.649	&	0.008	&	7.514	&	0.012 \\
5610.016	&	11.991	&	0.011	&	10.213	&	0.012	&	8.695	&	0.045	&	7.533	&	0.063 \\
5621.982	&	11.974	&	0.009	&	10.178	&	0.004	&	8.636	&	0.011	&	7.492	&	0.011 \\
5624.975	&	11.979	&	0.023	&	10.205	&	0.037	&	8.623	&	0.030	&	7.496	&	0.046 \\
5631.956	&	11.916	&	0.008	&	10.137	&	0.044	&	8.598	&	0.043	&	7.439	&	0.022 \\
5634.948	&	11.891	&	0.021	&	10.106	&	0.022	&	8.564	&	0.016	&	7.431	&	0.016 \\
5637.940	&	11.900	&	0.008	&	10.113	&	0.015	&	8.597	&	0.005	&	7.421	&	0.047 \\
5644.921	&	11.944	&	0.006	&	10.155	&	0.013	&	8.600	&	0.004	&	7.474	&	0.004 \\
5648.911	&	11.945	&	0.021	&	10.160	&	0.033	&	8.589	&	0.016	&	7.475	&	0.006 \\
5653.897	&	11.920	&	0.006	&	10.157	&	0.005	&	8.617	&	0.005	&	7.477	&	0.021 \\
5663.871	&	11.987	&	0.007	&	10.183	&	0.016	&	8.644	&	0.016	&	7.489	&	0.023 \\
5667.861	&	11.996	&	0.007	&	10.204	&	0.006	&	8.672	&	0.031	&	7.528	&	0.009 \\
5671.850	&	11.976	&	0.007	&	10.185	&	0.029	&	8.664	&	0.031	&	7.524	&	0.041 \\
5679.829	&	11.916	&	0.009	&	10.138	&	0.005	&	8.604	&	0.019	&	7.493	&	0.019 \\
5685.812	&	11.978	&	0.023	&	10.197	&	0.028	&	8.634	&	0.051	&	7.509	&	0.010 \\
5688.931	&	11.948	&	0.010	&	10.155	&	0.007	&	8.610	&	0.004	&	7.469	&	0.004 \\
5694.915	&	11.927	&	0.013	&	10.137	&	0.004	&	8.609	&	0.017	&	7.457	&	0.019 \\
5697.781	&	11.872	&	0.008	&	10.108	&	0.005	&	8.551	&	0.010	&	7.452	&	0.004 \\
5712.845	&	11.885	&	0.006	&	10.113	&	0.009	&	8.584	&	0.012	&	7.446	&	0.008 \\
5721.891	&	11.988	&	0.043	&	10.190	&	0.022	&	8.646	&	0.010	&	7.516	&	0.006 \\
5725.822	&	11.996	&	0.031	&	10.226	&	0.026	&	8.665	&	0.020	&	7.519	&	0.010 \\
5728.857	&	11.977	&	0.016	&	10.199	&	0.023	&	8.665	&	0.027	&	7.531	&	0.005 \\
5735.852	&	11.921	&	0.013	&	10.136	&	0.014	&	8.609	&	0.004	&	7.483	&	0.006 \\
5738.824	&	11.958	&	0.027	&	10.173	&	0.005	&	8.634	&	0.015	&	7.478	&	0.017 \\
5741.951	&	11.945	&	0.022	&	10.173	&	0.008	&	8.607	&	0.007	&	7.491	&	0.016 \\
5831.698	&	11.934	&	0.006	&	10.155	&	0.014	&	8.620	&	0.019	&	7.494	&	0.004 \\
5835.692	&	11.944	&	0.008	&	10.165	&	0.031	&	8.616	&	0.004	&	7.518	&	0.028 \\
5844.635	&	11.980	&	0.006	&	10.193	&	0.013	&	8.653	&	0.015	&	7.522	&	0.006 \\
5850.630	&	11.962	&	0.045	&	10.197	&	0.008	&	8.650	&	0.005	&	7.524	&	0.004 \\
5853.668	&	11.948	&	0.006	&	10.167	&	0.016	&	8.628	&	0.008	&	7.502	&	0.004 \\
5858.684	&	11.897	&	0.008	&	10.127	&	0.009	&	8.585	&	0.004	&	7.477	&	0.004 \\
5863.601	&	11.929	&	0.006	&	10.139	&	0.010	&	8.607	&	0.004	&	7.482	&	0.007 \\
5867.600	&	12.005	&	0.007	&	10.202	&	0.005	&	8.661	&	0.027	&	7.520	&	0.007 \\
5875.584	&	11.893	&	0.018	&	10.135	&	0.005	&	8.605	&	0.021	&	7.466	&	0.004 \\
5883.601	&	11.882	&	0.006	&	10.117	&	0.005	&	8.567	&	0.005	&	7.452	&	0.009 \\
6002.003	&	11.901	&	0.006	&	10.126	&	0.005	&	8.597	&	0.005	&	7.436	&	0.004 \\
6035.987	&	11.979	&	0.023	&	10.180	&	0.020	&	8.645	&	0.004	&	7.503	&	0.024 \\
\hline
\end{tabular}
\end{minipage}
\end{table*}

\begin{table*}
\begin{minipage}{160mm}
\contcaption{Photometric Measurements \label{table-phot}}
\begin{tabular}{lllllllll}
\hline

JD - 2,450,000		&	$B$		&	$\sigma_B$ &	$V$		&$\sigma_V$ 	&	$R$		& $\sigma_R$	&	$I$		& $\sigma_I$ \\ 
\hline
6039.979	&	11.945	&	0.006	&	10.155	&	0.005	&	8.624	&	0.008	&	7.486	&	0.009 \\
6042.906	&	11.934	&	0.006	&	10.136	&	0.009	&	8.609	&	0.011	&	7.463	&	0.008 \\
6045.979	&	11.947	&	0.012	&	10.158	&	0.005	&	8.628	&	0.005	&	7.480	&	0.004 \\
6048.878	&	11.989	&	0.016	&	10.211	&	0.015	&	8.651	&	0.039	&	7.514	&	0.038 \\
6052.941	&	11.953	&	0.015	&	10.174	&	0.005	&	8.633	&	0.004	&	7.499	&	0.017 \\
6067.970	&	11.899	&	0.011	&	10.118	&	0.006	&	8.588	&	0.010	&	7.457	&	0.033 \\
6070.836	&	11.866	&	0.017	&	10.090	&	0.006	&	8.566	&	0.018	&	7.427	&	0.005 \\
6073.836	&	11.919	&	0.008	&	10.138	&	0.038	&	8.609	&	0.009	&	7.449	&	0.019 \\
6076.864	&	11.898	&	0.019	&	10.120	&	0.017	&	8.560	&	0.036	&	7.442	&	0.027 \\
6086.868	&	11.942	&	0.007	&	10.155	&	0.018	&	8.640	&	0.027	&	7.484	&	0.005 \\
6089.838	&	11.959	&	0.017	&	10.180	&	0.032	&	8.652	&	0.017	&	7.481	&	0.011 \\
6092.845	&	11.987	&	0.008	&	10.180	&	0.005	&	8.650	&	0.005	&	7.507	&	0.005 \\
6097.826	&	11.982	&	0.016	&	10.193	&	0.005	&	8.655	&	0.005	&	7.503	&	0.005 \\
6104.847	&	11.937	&	0.009	&	10.123	&	0.008	&	8.596	&	0.004	&	7.460	&	0.019 \\
6190.733	&	11.933	&	0.030	&	10.158	&	0.006	&	8.627	&	0.004	&	7.489	&	0.012 \\
6193.640	&	11.966	&	0.022	&	10.166	&	0.004	&	8.630	&	0.011	&	7.489	&	0.013 \\
6202.693	&	11.975	&	0.012	&	10.176	&	0.005	&	8.660	&	0.022	&	7.526	&	0.004 \\
6205.610	&	11.964	&	0.009	&	10.169	&	0.011	&	8.646	&	0.004	&	7.515	&	0.008 \\
\hline
\end{tabular}
\end{minipage}
\end{table*}

\begin{table*}
\centering
\begin{minipage}{160mm}
\caption{$V^2$ Measurements from CHARA \label{table-vis2}}
\begin{tabular}{lrcccc}
\hline
       HJD - 2,450,000	&	    Baseline 	&	      $u$	&	      $v$ & $V^2$ & $\sigma_{V^2}$	\\
      (d)			&		(m)		&		(10$^6$ cycles radian$^{-1}$) & (10$^6$ cycles radian$^{-1}$) & & \\  \hline

       5442.6938 & 232.508 & 65.653 & 223.046 & 0.552 & 0.043 \\
       5443.7053 & 171.935 & -77.714 & 153.369 & 0.706 & 0.130 \\
       5443.7152 & 173.105 & -82.696 & 152.075 & 0.622 & 0.079 \\
       5443.7943 & 92.0942 & 91.829 & 6.9727 & 1.096 & 0.085 \\
       5443.8059 & 88.1392 & 87.721 & 8.5707 & 1.026 & 0.165 \\
       5459.7630 & 46.8597 & -21.801 & -41.479 & 1.297 & 0.059 \\
       5459.7695 & 45.9541 & -19.270 & -41.718 & 1.216 & 0.239 \\
       5460.6517 & 62.3953 & -51.795 & -34.790 & 0.869 & 0.053 \\
       5460.6600 & 61.4734 & -50.210 & -35.467 & 0.959 & 0.048 \\
       5460.6741 & 59.7343 & -47.242 & -36.556 & 0.874 & 0.054 \\
       5828.6275 & 239.741 & -171.085 & 167.945 & 0.415 & 0.074 \\
       5828.6275 & 312.117 & 299.632 & 87.394 & 0.573 & 0.052 \\
       5828.6275 & 285.871 & -128.547 & -255.339 & 0.427 & 0.097 \\
       5828.6472 & 244.850 & -183.352 & 162.276 & 0.471 & 0.173 \\
       5828.6472 & 306.665 & 290.973 & 96.840 & 0.505 & 0.100 \\
       5828.6472 & 280.577 & -107.621 & -259.116 & 0.439 & 0.067 \\
       5828.6657 & 248.009 & -192.273 & 156.651 & 0.454 & 0.075 \\
       5828.6657 & 298.036 & 278.787 & 105.372 & 0.527 & 0.122 \\
       5828.6657 & 275.937 & -86.514 & -262.023 & 0.454 & 0.125 \\
       5828.6860 & 249.381 & -199.063 & 150.217 & 0.378 & 0.045 \\
       5828.6860 & 284.970 & 261.066 & 114.248 & 0.596 & 0.047 \\
       5828.6860 & 271.636 & -62.003 & -264.465 & 0.421 & 0.044 \\
       5842.6526 & 249.360 & -200.189 & 148.677 & 0.579 & 0.082 \\
       5842.6526 & 281.406 & 256.275 & 116.243 & 0.611 & 0.050 \\
       5842.6526 & 270.793 & -56.086 & -264.921 & 0.508 & 0.072 \\
       5842.6636 & 248.785 & -202.104 & 145.079 & 0.620 & 0.084 \\
       5842.6636 & 272.484 & 244.283 & 120.720 & 0.782 & 0.053 \\
       5842.6636 & 269.125 & -42.178 & -265.800 & 0.690 & 0.084 \\
       5842.6765 & 247.165 & -203.101 & 140.857 & 0.423 & 0.057 \\
       5842.6765 & 261.055 & 228.827 & 125.650 & 0.416 & 0.134 \\
       5842.6765 & 267.746 & -25.726 & -266.507 & 0.526 & 0.066 \\
       6113.9098 & 299.267 & 46.283 & 295.666 & 0.564 & 0.049 \\
       6113.9224 & 297.775 & 27.543 & 296.499 & 0.706 & 0.219 \\
       6113.9357 & 297.049 & 9.5499 & 296.896 & 0.637 & 0.053 \\
       6113.9644 & 297.802 & -27.985 & 296.484 & 0.553 & 0.098 \\
       6113.9897 & 301.286 & -63.571 & 294.503 & 0.546 & 0.072 \\
       6114.0032 & 304.142 & -82.374 & 292.775 & 0.731 & 0.049 \\

\hline
\end{tabular}
\end{minipage}
\end{table*}

\begin{table*}
\begin{minipage}{160mm}
\contcaption{$V^2$ Measurements from CHARA \label{table-vis2}}
\begin{tabular}{lrcccc}
\hline
       HJD - 2,450,000	&	    Baseline	&	      $u$	&	      $v$ & $V^2$ & $\sigma_{V^2}$	\\
      (d)			&		(m)		&		(10$^6$ cycles radian$^{-1}$) & (10$^6$ cycles radian$^{-1}$) & & \\  \hline
       6115.8049 & 253.758 & -143.509 & 209.280 & 0.502 & 0.050 \\
       6115.8049 & 311.297 & 303.419 & 69.592 & 0.427 & 0.044 \\
       6115.8049 & 321.467 & -159.909 & -278.873 & 0.567 & 0.079 \\
       6115.8205 & 259.808 & -159.041 & 205.441 & 0.567 & 0.076 \\
       6115.8205 & 313.415 & 303.733 & 77.297 & 0.432 & 0.063 \\
       6115.8205 & 317.611 & -144.691 & -282.738 & 0.524 & 0.096 \\
       6140.7364 & 253.672 & -143.281 & 209.331 & 0.458 & 0.096 \\
       6140.7364 & 311.248 & 303.393 & 69.484 & 0.440 & 0.101 \\
       6140.7364 & 321.518 & -160.111 & -278.816 & 0.354 & 0.045 \\
       6140.7525 & 259.899 & -159.272 & 205.378 & 0.475 & 0.098 \\
       6140.7525 & 313.426 & 303.715 & 77.417 & 0.572 & 0.119 \\
       6140.7525 & 317.548 & -144.443 & -282.795 & 0.516 & 0.047 \\
       6140.7679 & 265.385 & -173.021 & 201.228 & 0.340 & 0.075 \\
       6140.7679 & 312.850 & 301.090 & 84.971 & 0.490 & 0.101 \\
       6140.7679 & 313.546 & -128.068 & -286.199 & 0.291 & 0.046 \\
       6140.7903 & 272.028 & -190.032 & 194.645 & 0.446 & 0.048 \\
       6140.7903 & 307.525 & 292.246 & 95.729 & 0.401 & 0.045 \\
       6140.7903 & 307.839 & -102.213 & -290.374 & 0.444 & 0.092 \\
       6140.8049 & 275.251 & -199.105 & 190.053 & 0.437 & 0.047 \\
       6140.8049 & 301.327 & 283.350 & 102.522 & 0.632 & 0.296 \\
       6140.8049 & 304.462 & -84.244 & -292.575 & 0.433 & 0.114 \\
       6185.6400 & 263.669 & -168.728 & 202.613 & 0.410 & 0.048 \\
       6185.6400 & 313.319 & 302.256 & 82.525 & 0.456 & 0.151 \\
       6185.6400 & 314.855 & -133.527 & -285.139 & 0.362 & 0.075 \\

       6185.6541 & 268.295 & -180.351 & 198.634 & 0.383 & 0.059 \\
       6185.6541 & 311.321 & 298.218 & 89.369 & 0.396 & 0.154 \\
       6185.6541 & 311.189 & -117.866 & -288.004 & 0.320 & 0.054 \\
       6185.6652 & 271.464 & -188.534 & 195.313 & 0.395 & 0.047 \\
       6185.6652 & 308.272 & 293.367 & 94.695 & 0.411 & 0.051 \\
       6185.6652 & 308.375 & -104.832 & -290.009 & 0.419 & 0.090 \\
       6185.6794 & 274.749 & -197.609 & 190.886 & 0.445 & 0.086 \\
       6185.6794 & 302.583 & 285.112 & 101.327 & 0.506 & 0.074 \\
       6185.6794 & 305.034 & -87.502 & -292.214 & 0.896 & 0.346 \\
       6185.6939 & 277.109 & -205.261 & 186.165 & 0.499 & 0.141 \\
       6185.6939 & 294.781 & 274.330 & 107.884 & 0.495 & 0.090 \\
       6185.6939 & 302.052 & -69.068 & -294.049 & 0.519 & 0.110 \\
       6516.6622 & 275.739 & 135.758 & 240.003 & 0.685 & 0.078 \\
       6516.6923 & 271.208 & 113.641 & 246.250 & 0.592 & 0.068 \\
       6516.6923 & 243.414 & -241.071 & -33.697 & 0.943 & 0.208 \\
       6516.6923 & 247.825 & 127.429 & -212.553 & 0.666 & 0.060 \\
       6516.7182 & 266.829 & 91.736 & 250.564 & 0.534 & 0.076 \\
       6516.7182 & 250.177 & -246.290 & -43.932 & 0.382 & 0.233 \\
       6516.7182 & 258.037 & 154.553 & -206.631 & 0.560 & 0.097 \\
       6517.6701 & 274.233 & 128.202 & 242.421 & 0.684 & 0.130 \\
       6517.6701 & 234.346 & -232.877 & -26.192 & 0.594 & 0.203 \\
       6517.6701 & 240.233 & 104.675 & -216.229 & 0.695 & 0.101 \\
       6517.6893 & 271.252 & 113.856 & 246.200 & 0.845 & 0.080 \\
       6517.6893 & 243.311 & -240.982 & -33.591 & 0.796 & 0.153 \\
       6517.6893 & 247.717 & 127.125 & -212.609 & 0.390 & 0.087 \\
       6565.6292 & 272.530 & -191.380 & 194.027 & 0.510 & 0.059 \\
       6565.6292 & 306.793 & 291.163 & 96.677 & 0.568 & 0.072 \\
       6565.6292 & 307.352 & -99.782 & -290.704 & 0.517 & 0.053 \\
       6565.6431 & 275.502 & -199.870 & 189.613 & 0.453 & 0.158 \\
       6565.6431 & 300.639 & 282.391 & 103.146 & 0.867 & 0.217 \\
       6565.6431 & 304.168 & -82.520 & -292.760 & 0.459 & 0.168 \\
\hline

\end{tabular}
\end{minipage}
\end{table*}

\begin{table*}
\centering
\begin{minipage}{160mm}
\caption{Orbital light curve from {\it MOST}}
\begin{tabular}{lllllllll}
\hline
JD 			&	    Phase 	&	    $\triangle m$	&	 JD 	&	    Phase 	&	    $\triangle m$	& JD 	&	    Phase 	&	    $\triangle m$	  	\\
- 2,450,000	& 			&	(mag)	& - 2,450,000	& 			&	(mag)	& - 2,450,000	& 			&	(mag)	\\ \hline

6827.5752	&	0.9074	&	0.00820	&	6832.1544	&	0.9828	&	-0.00442	&	6836.3796	&	0.0524	&	0.00816	\\
6827.6452	&	0.9085	&	0.00802	&	6832.2234	&	0.9839	&	0.00577	&	6836.4493	&	0.0535	&	0.00706	\\
6827.7156	&	0.9097	&	0.00795	&	6832.2936	&	0.9851	&	0.00772	&	6836.5224	&	0.0547	&	0.01426	\\
6827.7864	&	0.9109	&	0.01235	&	6832.3650	&	0.9863	&	-0.01125	&	6836.5842	&	0.0557	&	0.01600	\\
6827.8569	&	0.9120	&	0.01427	&	6832.4349	&	0.9874	&	-0.01242	&	6836.6535	&	0.0569	&	0.02378	\\
6827.9273	&	0.9132	&	0.01204	&	6832.5051	&	0.9886	&	-0.01085	&	6836.7230	&	0.0580	&	0.02525	\\
6827.9972	&	0.9143	&	0.01216	&	6832.5768	&	0.9897	&	-0.00125	&	6836.7947	&	0.0592	&	0.03187	\\
6828.0684	&	0.9155	&	0.01508	&	6832.6469	&	0.9909	&	-0.00493	&	6837.7800	&	0.0754	&	0.04240	\\
6828.1389	&	0.9167	&	0.01289	&	6832.7175	&	0.9921	&	-0.00373	&	6839.2788	&	0.1001	&	0.05623	\\
6828.2090	&	0.9178	&	0.01174	&	6832.7881	&	0.9932	&	-0.00160	&	6839.3313	&	0.1010	&	0.05782	\\
6828.2796	&	0.9190	&	0.01143	&	6832.8581	&	0.9944	&	-0.00331	&	6839.4048	&	0.1022	&	0.05720	\\
6828.3488	&	0.9201	&	0.00061	&	6832.9286	&	0.9955	&	-0.00575	&	6839.4737	&	0.1033	&	0.05796	\\
6828.4186	&	0.9213	&	-0.01838	&	6832.9994	&	0.9967	&	-0.00193	&	6839.5459	&	0.1045	&	0.05643	\\
6828.8432	&	0.9283	&	0.00079	&	6833.0697	&	0.9979	&	0.00270	&	6839.6122	&	0.1056	&	0.05788	\\
6828.9134	&	0.9294	&	0.00207	&	6833.1397	&	0.9990	&	-0.00520	&	6839.6730	&	0.1066	&	0.05915	\\
6828.9840	&	0.9306	&	0.00721	&	6833.2104	&	0.0002	&	-0.00190	&	6839.7473	&	0.1078	&	0.06494	\\
6829.0543	&	0.9317	&	0.00774	&	6833.2799	&	0.0013	&	-0.00538	&	6839.8159	&	0.1089	&	0.06572	\\
6829.1246	&	0.9329	&	0.00416	&	6833.3503	&	0.0025	&	0.00320	&	6839.8878	&	0.1101	&	0.06329	\\
6829.1950	&	0.9341	&	0.00503	&	6833.4208	&	0.0036	&	0.00924	&	6839.9577	&	0.1113	&	0.05945	\\
6829.2653	&	0.9352	&	0.03376	&	6833.4912	&	0.0048	&	-0.00932	&	6840.0267	&	0.1124	&	0.05959	\\
6829.3345	&	0.9364	&	0.00897	&	6833.5628	&	0.0060	&	-0.00268	&	6840.0977	&	0.1136	&	0.05971	\\
6829.4058	&	0.9375	&	-0.00834	&	6833.6335	&	0.0071	&	0.00216	&	6840.1957	&	0.1152	&	0.05303	\\
6829.4761	&	0.9387	&	-0.00641	&	6833.7033	&	0.0083	&	-0.00413	&	6840.2449	&	0.1160	&	0.05419	\\
6829.5472	&	0.9399	&	0.00039	&	6833.7738	&	0.0094	&	-0.00021	&	6840.3181	&	0.1172	&	0.05148	\\
6829.6183	&	0.9410	&	-0.00026	&	6833.8445	&	0.0106	&	-0.00216	&	6840.3913	&	0.1184	&	0.05015	\\
6829.6879	&	0.9422	&	0.00299	&	6833.9145	&	0.0118	&	0.00564	&	6840.4617	&	0.1196	&	0.05459	\\
6829.7591	&	0.9433	&	-0.00371	&	6833.9854	&	0.0129	&	0.00338	&	6840.5291	&	0.1207	&	0.05575	\\
6829.8295	&	0.9445	&	0.00357	&	6834.0554	&	0.0141	&	0.00818	&	6840.6001	&	0.1218	&	0.06014	\\
6829.8999	&	0.9457	&	-0.00823	&	6834.1259	&	0.0152	&	0.00662	&	6840.6713	&	0.1230	&	0.06456	\\
6829.9704	&	0.9468	&	0.00077	&	6834.1963	&	0.0164	&	0.01064	&	6840.7469	&	0.1243	&	0.07149	\\
6830.0410	&	0.9480	&	-0.00401	&	6834.2662	&	0.0176	&	0.00068	&	6840.8175	&	0.1254	&	0.06395	\\
6830.1112	&	0.9491	&	-0.00381	&	6834.3364	&	0.0187	&	-0.00743	&	6840.8864	&	0.1266	&	0.05920	\\
6830.1813	&	0.9503	&	-0.00519	&	6834.4063	&	0.0199	&	-0.02639	&	6840.9562	&	0.1277	&	0.05512	\\
6830.2506	&	0.9514	&	0.00439	&	6834.4777	&	0.0210	&	-0.00397	&	6841.0196	&	0.1287	&	0.05563	\\
6830.3212	&	0.9526	&	-0.00316	&	6834.5490	&	0.0222	&	-0.00527	&	6841.0959	&	0.1300	&	0.05677	\\
6830.3917	&	0.9538	&	-0.02045	&	6834.6203	&	0.0234	&	-0.00009	&	6841.1639	&	0.1311	&	0.05555	\\
6830.4620	&	0.9549	&	-0.02464	&	6834.6896	&	0.0245	&	-0.00444	&	6841.2356	&	0.1323	&	0.05255	\\
6830.5339	&	0.9561	&	-0.00531	&	6834.7608	&	0.0257	&	0.00046	&	6841.3054	&	0.1335	&	0.05054	\\
6830.6042	&	0.9573	&	-0.00388	&	6834.8301	&	0.0268	&	-0.00115	&	6841.3755	&	0.1346	&	0.05461	\\
6830.6750	&	0.9584	&	-0.00823	&	6834.9013	&	0.0280	&	0.00818	&	6841.4481	&	0.1358	&	0.06025	\\
6830.7454	&	0.9596	&	-0.01100	&	6834.9716	&	0.0292	&	0.00543	&	6841.5153	&	0.1369	&	0.05872	\\
6830.8160	&	0.9607	&	-0.00541	&	6835.0420	&	0.0303	&	0.01480	&	6841.5879	&	0.1381	&	0.05827	\\
6830.8863	&	0.9619	&	-0.00278	&	6835.1125	&	0.0315	&	0.00742	&	6841.6581	&	0.1393	&	0.05432	\\
6830.9569	&	0.9631	&	-0.00594	&	6835.1829	&	0.0326	&	0.01615	&	6841.7287	&	0.1404	&	0.04616	\\
6831.0272	&	0.9642	&	-0.00492	&	6835.2525	&	0.0338	&	0.02036	&	6841.8027	&	0.1416	&	0.03994	\\
6831.0973	&	0.9654	&	-0.01175	&	6835.3225	&	0.0349	&	0.00389	&	6841.8702	&	0.1428	&	0.03817	\\
6831.1679	&	0.9665	&	-0.00384	&	6835.3934	&	0.0361	&	-0.00833	&	6841.9394	&	0.1439	&	0.03800	\\
6831.2370	&	0.9677	&	0.00352	&	6835.4646	&	0.0373	&	-0.00295	&	6842.0052	&	0.1450	&	0.04399	\\
6831.3072	&	0.9688	&	-0.01404	&	6835.5353	&	0.0385	&	0.00110	&	6842.0801	&	0.1462	&	0.04466	\\
6831.3782	&	0.9700	&	-0.01552	&	6835.6059	&	0.0396	&	0.01702	&	6842.1465	&	0.1473	&	0.04831	\\
6831.4479	&	0.9712	&	-0.03032	&	6835.6762	&	0.0408	&	0.00319	&	6842.2214	&	0.1485	&	0.04867	\\
6831.5203	&	0.9723	&	-0.02038	&	6835.7470	&	0.0419	&	0.00620	&	6842.2912	&	0.1497	&	0.04700	\\
6831.5899	&	0.9735	&	-0.00677	&	6835.8174	&	0.0431	&	0.01017	&	6842.3614	&	0.1508	&	0.04522	\\
6831.6611	&	0.9747	&	-0.01306	&	6835.8871	&	0.0442	&	0.00591	&	6842.4337	&	0.1520	&	0.04290	\\
6831.7315	&	0.9758	&	-0.01440	&	6835.9575	&	0.0454	&	0.00987	&	6842.4995	&	0.1531	&	0.03809	\\
6831.8017	&	0.9770	&	-0.00947	&	6836.0280	&	0.0466	&	0.02004	&	6842.5727	&	0.1543	&	0.03615	\\
6831.8720	&	0.9781	&	-0.00791	&	6836.0988	&	0.0477	&	0.01839	&	6842.6434	&	0.1555	&	0.03611	\\
6831.9429	&	0.9793	&	-0.00785	&	6836.1692	&	0.0489	&	0.01539	&	6842.7160	&	0.1567	&	0.03792	\\
6832.0128	&	0.9805	&	0.00241	&	6836.2383	&	0.0500	&	0.02371	&	6842.7858	&	0.1578	&	0.03724	\\
6832.0838	&	0.9816	&	-0.00308	&	6836.3089	&	0.0512	&	0.01848	&	6842.8556	&	0.1590	&	0.03884	\\

\hline

\end{tabular}
\end{minipage}
\end{table*}

\begin{table*}
\centering
\begin{minipage}{160mm}
\contcaption{Orbital light curve from {\it MOST}}
\begin{tabular}{lllllllll}
\hline
JD 			&	    Phase 	&	    $\triangle m$	&	 JD 	&	    Phase 	&	    $\triangle m$	& JD 	&	    Phase 	&	    $\triangle m$	  	\\
- 2,450,000	& 			&	(mag)	& - 2,450,000	& 			&	(mag)	& - 2,450,000	& 			&	(mag)	\\ \hline

6842.9280	&	0.1602	&	0.04261	&	6853.4190	&	0.3329	&	-0.04118	&	6857.6448	&	0.4025	&	-0.04528	\\
6842.9974	&	0.1613	&	0.04537	&	6853.4886	&	0.3341	&	-0.04578	&	6857.7144	&	0.4036	&	-0.04485	\\
6843.0648	&	0.1624	&	0.04647	&	6853.5588	&	0.3352	&	-0.04650	&	6857.7631	&	0.4044	&	-0.04807	\\
6843.1376	&	0.1636	&	0.04730	&	6853.6298	&	0.3364	&	-0.04809	&	6859.3455	&	0.4305	&	-0.03419	\\
6843.2088	&	0.1648	&	0.04153	&	6853.7017	&	0.3376	&	-0.04638	&	6859.4002	&	0.4314	&	-0.02815	\\
6843.2718	&	0.1658	&	0.04231	&	6853.7698	&	0.3387	&	-0.04539	&	6859.4605	&	0.4324	&	-0.02368	\\
6843.3470	&	0.1671	&	0.03855	&	6853.8429	&	0.3399	&	-0.04695	&	6860.2651	&	0.4456	&	-0.02287	\\
6843.4217	&	0.1683	&	0.03433	&	6853.9112	&	0.3410	&	-0.04553	&	6860.3200	&	0.4465	&	-0.03143	\\
6843.4888	&	0.1694	&	0.04293	&	6853.9821	&	0.3422	&	-0.04005	&	6860.3902	&	0.4477	&	-0.02512	\\
6843.5608	&	0.1706	&	0.04146	&	6854.0531	&	0.3433	&	-0.03356	&	6860.4603	&	0.4488	&	-0.03027	\\
6843.6319	&	0.1718	&	0.04570	&	6854.1245	&	0.3445	&	-0.03652	&	6860.5307	&	0.4500	&	-0.02811	\\
6843.6992	&	0.1729	&	0.05073	&	6854.1916	&	0.3456	&	-0.04175	&	6860.6021	&	0.4512	&	-0.03394	\\
6843.7721	&	0.1741	&	0.04827	&	6854.2627	&	0.3468	&	-0.04285	&	6860.6676	&	0.4523	&	-0.01720	\\
6843.8430	&	0.1752	&	0.04627	&	6854.3327	&	0.3479	&	-0.04590	&	6860.7458	&	0.4535	&	-0.02614	\\
6843.9139	&	0.1764	&	0.04000	&	6854.4037	&	0.3491	&	-0.04148	&	6860.8145	&	0.4547	&	-0.02307	\\
6843.9826	&	0.1775	&	0.03762	&	6854.4739	&	0.3503	&	-0.03919	&	6860.8823	&	0.4558	&	-0.02226	\\
6844.0540	&	0.1787	&	0.03054	&	6854.5450	&	0.3514	&	-0.03659	&	6860.9521	&	0.4569	&	-0.02177	\\
6844.1216	&	0.1798	&	0.02976	&	6854.6157	&	0.3526	&	-0.03042	&	6861.0251	&	0.4581	&	-0.02052	\\
6844.1918	&	0.1810	&	0.02832	&	6854.6904	&	0.3538	&	-0.02877	&	6861.0917	&	0.4592	&	-0.01903	\\
6844.2643	&	0.1822	&	0.02399	&	6854.7566	&	0.3549	&	-0.03074	&	6861.1581	&	0.4603	&	-0.01971	\\
6844.3347	&	0.1833	&	0.02292	&	6854.8302	&	0.3561	&	-0.03000	&	6861.2353	&	0.4616	&	-0.02122	\\
6844.4031	&	0.1845	&	0.02804	&	6854.8973	&	0.3572	&	-0.03682	&	6861.3063	&	0.4628	&	-0.02858	\\
6844.4734	&	0.1856	&	0.02990	&	6854.9685	&	0.3584	&	-0.04125	&	6861.3756	&	0.4639	&	-0.02976	\\
6844.5468	&	0.1868	&	0.03491	&	6855.0402	&	0.3596	&	-0.04338	&	6861.4458	&	0.4651	&	-0.02878	\\
6844.6183	&	0.1880	&	0.03654	&	6855.1073	&	0.3607	&	-0.04329	&	6861.5165	&	0.4662	&	-0.02567	\\
6844.6873	&	0.1891	&	0.03621	&	6855.1778	&	0.3619	&	-0.04182	&	6861.5898	&	0.4674	&	-0.02274	\\
6844.7518	&	0.1902	&	0.03674	&	6855.2470	&	0.3630	&	-0.04142	&	6861.6597	&	0.4686	&	-0.02256	\\
6844.8255	&	0.1914	&	0.03653	&	6855.3203	&	0.3642	&	-0.03787	&	6861.7303	&	0.4697	&	-0.01747	\\
6844.8770	&	0.1923	&	0.04554	&	6855.3893	&	0.3653	&	-0.03743	&	6861.7984	&	0.4709	&	-0.02051	\\
6848.0850	&	0.2451	&	0.00048	&	6855.4637	&	0.3666	&	-0.03473	&	6861.8696	&	0.4720	&	-0.01779	\\
6848.1374	&	0.2459	&	0.00382	&	6855.5323	&	0.3677	&	-0.03460	&	6861.9403	&	0.4732	&	-0.00595	\\
6848.2073	&	0.2471	&	0.00014	&	6855.6028	&	0.3689	&	-0.03850	&	6862.0069	&	0.4743	&	-0.01727	\\
6848.2781	&	0.2483	&	-0.00217	&	6855.6730	&	0.3700	&	-0.03758	&	6862.0786	&	0.4755	&	-0.01329	\\
6848.3492	&	0.2494	&	-0.00336	&	6855.7461	&	0.3712	&	-0.04064	&	6862.1474	&	0.4766	&	-0.00540	\\
6848.4193	&	0.2506	&	0.00030	&	6855.8160	&	0.3724	&	-0.03970	&	6862.2134	&	0.4777	&	0.00300	\\
6848.4888	&	0.2517	&	-0.00276	&	6855.8822	&	0.3735	&	-0.04297	&	6862.2908	&	0.4790	&	0.01150	\\
6848.5547	&	0.2528	&	-0.00657	&	6855.9546	&	0.3747	&	-0.04321	&	6862.3347	&	0.4797	&	-0.02169	\\
6848.6287	&	0.2540	&	-0.01074	&	6856.0262	&	0.3758	&	-0.04070	&	6862.4513	&	0.4816	&	-0.02406	\\
6848.7024	&	0.2552	&	-0.01152	&	6856.0900	&	0.3769	&	-0.03539	&	6862.5026	&	0.4825	&	-0.01163	\\
6848.7710	&	0.2564	&	-0.00968	&	6856.1627	&	0.3781	&	-0.02474	&	6862.5726	&	0.4836	&	-0.00850	\\
6848.8455	&	0.2576	&	-0.01502	&	6856.2346	&	0.3793	&	-0.02727	&	6862.6445	&	0.4848	&	-0.01311	\\
6848.9116	&	0.2587	&	-0.01942	&	6856.3061	&	0.3804	&	-0.02991	&	6862.7123	&	0.4859	&	-0.01690	\\
6848.9825	&	0.2599	&	-0.01607	&	6856.3751	&	0.3816	&	-0.02864	&	6862.7812	&	0.4871	&	-0.01334	\\
6849.0499	&	0.2610	&	-0.01132	&	6856.4481	&	0.3828	&	-0.03227	&	6862.8536	&	0.4882	&	-0.01190	\\
6849.1208	&	0.2621	&	-0.00555	&	6856.5183	&	0.3839	&	-0.03410	&	6862.9214	&	0.4894	&	-0.02091	\\
6849.1943	&	0.2633	&	-0.00257	&	6856.5871	&	0.3851	&	-0.03461	&	6862.9938	&	0.4906	&	-0.02148	\\
6849.2659	&	0.2645	&	-0.00270	&	6856.6608	&	0.3863	&	-0.03427	&	6863.0655	&	0.4917	&	-0.02445	\\
6849.3328	&	0.2656	&	-0.00615	&	6856.7267	&	0.3874	&	-0.03557	&	6863.1326	&	0.4928	&	-0.02499	\\
6849.4054	&	0.2668	&	-0.00636	&	6856.8022	&	0.3886	&	-0.03091	&	6863.1995	&	0.4939	&	-0.01970	\\
6849.4553	&	0.2676	&	-0.01642	&	6856.8738	&	0.3898	&	-0.03651	&	6863.2755	&	0.4952	&	-0.01856	\\
6852.7259	&	0.3215	&	-0.04401	&	6856.9380	&	0.3908	&	-0.03627	&	6863.3467	&	0.4964	&	-0.01178	\\
6852.7890	&	0.3225	&	-0.04841	&	6857.0092	&	0.3920	&	-0.04155	&	6863.4177	&	0.4975	&	-0.01084	\\
6852.8552	&	0.3236	&	-0.05120	&	6857.0819	&	0.3932	&	-0.03586	&	6863.4821	&	0.4986	&	-0.01188	\\
6852.9265	&	0.3248	&	-0.04577	&	6857.1494	&	0.3943	&	-0.03580	&	6863.5579	&	0.4998	&	-0.01291	\\
6852.9962	&	0.3259	&	-0.04957	&	6857.2187	&	0.3955	&	-0.04584	&	6863.6281	&	0.5010	&	-0.02005	\\
6853.0633	&	0.3270	&	-0.03773	&	6857.2905	&	0.3966	&	-0.04543	&	6863.6998	&	0.5022	&	-0.02510	\\
6853.1344	&	0.3282	&	-0.03876	&	6857.3570	&	0.3977	&	-0.04551	&	6863.7691	&	0.5033	&	-0.02042	\\
6853.2075	&	0.3294	&	-0.03760	&	6857.4310	&	0.3990	&	-0.04588	&	6863.8413	&	0.5045	&	-0.01763	\\
6853.2736	&	0.3305	&	-0.03834	&	6857.5018	&	0.4001	&	-0.05396	&	6863.9123	&	0.5057	&	-0.01640	\\
6853.3438	&	0.3317	&	-0.04217	&	6857.5726	&	0.4013	&	-0.04652	&	6863.9811	&	0.5068	&	-0.02477	\\

\hline

\end{tabular}
\end{minipage}
\end{table*}

\begin{table*}
\centering
\begin{minipage}{160mm}
\contcaption{Orbital light curve from {\it MOST}}
\begin{tabular}{lllllllll}
\hline
JD 			&	    Phase 	&	    $\triangle m$	&	 JD 	&	    Phase 	&	    $\triangle m$	& JD 	&	    Phase 	&	    $\triangle m$	  	\\
- 2,450,000	& 			&	(mag)	& - 2,450,000	& 			&	(mag)	& - 2,450,000	& 			&	(mag)	\\ \hline

6864.0514	&	0.5080	&	-0.02341	&	6864.8261	&	0.5207	&	-0.01528	&	6868.8420	&	0.5868	&	-0.00514	\\
6864.1220	&	0.5091	&	-0.02047	&	6864.8986	&	0.5219	&	-0.01304	&	6868.9894	&	0.5893	&	0.00029	\\
6864.1923	&	0.5103	&	-0.01894	&	6864.9685	&	0.5231	&	-0.01478	&	6869.0531	&	0.5903	&	-0.00131	\\
6864.2616	&	0.5114	&	-0.01585	&	6865.0374	&	0.5242	&	-0.01522	&	6869.1235	&	0.5915	&	0.00349	\\
6864.3325	&	0.5126	&	-0.01442	&	6865.1080	&	0.5254	&	-0.01815	&	6869.1928	&	0.5926	&	0.00137	\\
6864.4097	&	0.5139	&	-0.01857	&	6865.1787	&	0.5265	&	-0.02022	&	6869.2640	&	0.5938	&	0.00109	\\
6864.4738	&	0.5149	&	-0.02090	&	6865.2513	&	0.5277	&	-0.02008	&	6869.3326	&	0.5949	&	0.00496	\\
6864.5445	&	0.5161	&	-0.01994	&	6865.3190	&	0.5288	&	-0.02510	&	6869.4040	&	0.5961	&	0.00343	\\
6864.6154	&	0.5173	&	-0.02043	&	6865.3875	&	0.5300	&	-0.02227	&	6869.4738	&	0.5972	&	-0.00084	\\
6864.6861	&	0.5184	&	-0.02079	&	6865.4597	&	0.5312	&	-0.02362	&	6869.5458	&	0.5984	&	0.00061	\\
6864.7562	&	0.5196	&	-0.01275	&	6868.7799	&	0.5858	&	-0.00274	&	6869.6164	&	0.5996	&	0.00379	\\

\hline

\end{tabular}
\end{minipage}
\end{table*}

\begin{table*}
\centering
\begin{minipage}{160mm}
\caption{Pulsational light curve from {\it MOST}}
\begin{tabular}{llllllllll}
\hline
JD 	&  $\triangle m$		&	JD 	&  $\triangle m$		&JD 	&  $\triangle m$		&JD 	&  $\triangle m$		&JD 	&  $\triangle m$		\\
 - 2,450,000	& (mag)	&  - 2,450,000	& (mag)	&  - 2,450,000	& (mag)	&  - 2,450,000	& (mag)	&  - 2,450,000	& (mag)	\\ \hline
6827.5753	&	0.00733	&	6833.7763	&	0.00080	&	6841.0668	&	-0.00696	&	6848.5935	&	0.00229	&	6856.1150	&	0.00565	\\
6827.6459	&	0.00627	&	6833.8471	&	-0.00255	&	6841.1386	&	-0.00599	&	6848.6641	&	0.00016	&	6856.1864	&	0.00559	\\
6827.7164	&	0.00475	&	6833.9178	&	0.00068	&	6841.2086	&	-0.00566	&	6848.7312	&	-0.00070	&	6856.2571	&	0.00528	\\
6827.7871	&	0.00264	&	6833.9879	&	-0.00031	&	6841.2815	&	-0.00493	&	6848.8045	&	-0.00104	&	6856.3371	&	0.00376	\\
6827.8579	&	-0.00107	&	6834.0591	&	0.00398	&	6841.3531	&	0.00181	&	6848.8771	&	-0.01215	&	6856.4108	&	-0.00559	\\
6827.9291	&	-0.00801	&	6834.1295	&	-0.00384	&	6841.4261	&	0.00420	&	6848.9516	&	-0.00793	&	6856.4819	&	-0.01185	\\
6827.9993	&	-0.00739	&	6834.1999	&	-0.00202	&	6841.5038	&	0.00727	&	6849.0253	&	-0.00299	&	6856.5496	&	-0.00970	\\
6828.0702	&	-0.00465	&	6834.2705	&	-0.00280	&	6841.5779	&	0.00556	&	6849.0968	&	0.00220	&	6856.6229	&	-0.01001	\\
6828.1407	&	0.00171	&	6834.5490	&	0.00505	&	6841.6504	&	0.00285	&	6849.1676	&	0.00657	&	6856.6909	&	-0.00903	\\
6828.2112	&	0.00217	&	6834.6200	&	0.00799	&	6841.7268	&	-0.00124	&	6849.2372	&	0.00487	&	6856.7572	&	-0.00325	\\
6828.8436	&	-0.00236	&	6834.6914	&	0.00534	&	6841.7994	&	-0.00075	&	6849.3129	&	0.00545	&	6856.8293	&	0.00050	\\
6828.9147	&	-0.00340	&	6834.7619	&	0.00217	&	6841.8860	&	-0.00486	&	6849.3794	&	-0.00628	&	6856.9030	&	0.00763	\\
6828.9833	&	0.00407	&	6834.8326	&	-0.00218	&	6841.9566	&	-0.00275	&	6849.4509	&	-0.01807	&	6856.9746	&	0.00772	\\
6829.0562	&	0.00640	&	6834.9027	&	0.00004	&	6842.0321	&	0.00116	&	6852.7149	&	-0.00531	&	6857.0488	&	0.00771	\\
6829.1260	&	0.00702	&	6834.9738	&	-0.00553	&	6842.0975	&	0.00577	&	6852.7954	&	-0.00292	&	6857.1181	&	0.00316	\\
6829.1966	&	0.00335	&	6835.0446	&	-0.00427	&	6842.1701	&	0.00480	&	6852.8645	&	-0.00280	&	6857.1934	&	0.00564	\\
6829.2680	&	-0.00106	&	6835.1150	&	-0.00598	&	6842.2395	&	0.00813	&	6852.9366	&	0.00874	&	6857.2647	&	-0.00077	\\
6829.5469	&	-0.00729	&	6835.1855	&	0.00346	&	6842.3314	&	0.00144	&	6853.0177	&	0.01266	&	6857.3374	&	-0.00140	\\
6829.6184	&	-0.00695	&	6835.2560	&	0.00396	&	6842.4032	&	-0.00060	&	6853.0853	&	0.01653	&	6857.4078	&	-0.00372	\\
6829.6890	&	-0.00373	&	6835.5357	&	-0.00367	&	6842.4739	&	-0.01005	&	6853.1568	&	0.01550	&	6857.4782	&	-0.00556	\\
6829.7598	&	-0.00247	&	6835.6062	&	-0.00430	&	6842.5445	&	-0.01488	&	6853.2297	&	0.00970	&	6857.5505	&	-0.00543	\\
6829.8302	&	0.00404	&	6835.6773	&	-0.00640	&	6842.6167	&	-0.01600	&	6853.3105	&	0.00384	&	6857.6195	&	-0.00103	\\
6829.9010	&	0.00211	&	6835.7478	&	-0.00327	&	6842.6896	&	-0.01083	&	6853.3839	&	-0.00157	&	6857.6908	&	0.00393	\\
6829.9719	&	0.00494	&	6835.8185	&	0.00135	&	6842.7808	&	-0.00289	&	6853.4547	&	-0.00755	&	6857.7555	&	0.00769	\\
6830.0424	&	0.00493	&	6835.8892	&	0.00027	&	6842.8683	&	0.00348	&	6853.5242	&	-0.01434	&	6859.3218	&	0.00238	\\
6830.1126	&	0.00157	&	6835.9600	&	0.00556	&	6842.9433	&	0.01316	&	6853.6012	&	-0.01420	&	6859.3955	&	0.00381	\\
6830.1835	&	-0.00087	&	6836.0305	&	0.00515	&	6843.0161	&	0.01633	&	6853.6693	&	-0.00750	&	6859.4662	&	-0.00244	\\
6830.2539	&	-0.00246	&	6836.1009	&	0.01547	&	6843.0864	&	0.01684	&	6853.7378	&	-0.00300	&	6860.2527	&	0.00304	\\
6830.5364	&	0.00514	&	6836.1717	&	0.00518	&	6843.1610	&	0.00941	&	6853.8023	&	0.00057	&	6860.3217	&	0.00282	\\
6830.6074	&	0.00367	&	6836.2423	&	0.00325	&	6843.2301	&	0.00194	&	6853.8791	&	0.00484	&	6860.3942	&	-0.00648	\\
6830.6779	&	0.00334	&	6836.5189	&	-0.00476	&	6843.3076	&	-0.00004	&	6853.9507	&	0.01128	&	6860.4647	&	-0.00862	\\
6830.8159	&	-0.00072	&	6836.5911	&	-0.00554	&	6843.3820	&	-0.00756	&	6854.0261	&	0.01047	&	6860.5370	&	-0.00929	\\
6830.8869	&	-0.00205	&	6836.6619	&	-0.00122	&	6843.4499	&	-0.00826	&	6854.0945	&	0.00465	&	6860.6090	&	-0.00932	\\
6830.9573	&	-0.00148	&	6836.7327	&	0.00244	&	6843.5238	&	-0.00867	&	6854.1664	&	0.00007	&	6860.6822	&	-0.00525	\\
6831.0281	&	-0.00154	&	6836.8036	&	0.00625	&	6843.5910	&	-0.00520	&	6854.2368	&	-0.00085	&	6860.7529	&	0.00003	\\
6831.0988	&	-0.00507	&	6837.7828	&	0.00025	&	6843.6641	&	0.00218	&	6854.3083	&	-0.00360	&	6860.8235	&	0.00329	\\
6831.1690	&	0.00162	&	6839.2853	&	0.01004	&	6843.7290	&	0.00525	&	6854.3783	&	0.00308	&	6860.8957	&	0.00185	\\
6831.2395	&	0.00348	&	6839.3684	&	0.00679	&	6843.8025	&	0.00851	&	6854.4512	&	0.00127	&	6860.9660	&	0.00167	\\
6831.5232	&	0.01217	&	6839.4381	&	0.00014	&	6843.8757	&	0.00054	&	6854.5200	&	0.00242	&	6861.0372	&	0.00090	\\
6831.6611	&	0.00517	&	6839.5106	&	-0.00521	&	6843.9431	&	0.00197	&	6854.5920	&	0.00685	&	6861.1075	&	0.00067	\\
6831.7319	&	0.00076	&	6839.5819	&	-0.00929	&	6844.0165	&	0.00109	&	6854.6622	&	0.00531	&	6861.2299	&	-0.00161	\\
6831.8030	&	-0.00258	&	6839.6561	&	-0.00389	&	6844.0904	&	-0.00240	&	6854.7294	&	0.00886	&	6861.2968	&	0.00053	\\
6831.8734	&	-0.00557	&	6839.7552	&	0.00100	&	6844.1638	&	-0.00582	&	6854.8012	&	0.00414	&	6861.3698	&	0.00053	\\
6831.9443	&	-0.00854	&	6839.8306	&	0.01277	&	6844.2317	&	-0.00924	&	6854.8724	&	-0.00657	&	6861.4416	&	0.00665	\\
6832.0139	&	-0.00634	&	6839.8998	&	0.01235	&	6844.3031	&	-0.00627	&	6854.9465	&	-0.00423	&	6861.5175	&	0.00338	\\
6832.0856	&	-0.00227	&	6839.9708	&	0.01620	&	6844.3754	&	-0.00090	&	6855.0152	&	-0.01034	&	6861.5931	&	0.00524	\\
6832.1559	&	-0.00393	&	6840.0425	&	0.02147	&	6844.4499	&	0.00152	&	6855.0865	&	-0.01359	&	6861.6661	&	0.00380	\\
6832.2257	&	0.00273	&	6840.1136	&	0.01248	&	6844.5194	&	0.00469	&	6855.1610	&	-0.00659	&	6861.7380	&	-0.00078	\\
6832.5092	&	-0.00501	&	6840.1978	&	-0.00033	&	6844.5866	&	0.00557	&	6855.2307	&	-0.00122	&	6861.8085	&	-0.00258	\\
6832.5797	&	-0.00441	&	6840.2760	&	0.00270	&	6844.6559	&	0.00730	&	6855.2997	&	-0.00089	&	6861.8785	&	-0.00572	\\
6832.7193	&	-0.00375	&	6840.3509	&	-0.00265	&	6844.7259	&	0.00257	&	6855.3723	&	0.00487	&	6861.9901	&	-0.02406	\\
6832.7899	&	-0.00007	&	6840.4201	&	-0.00267	&	6844.7962	&	0.00081	&	6855.4443	&	0.00637	&	6862.0607	&	-0.01701	\\
6832.8608	&	0.00479	&	6840.4940	&	-0.00251	&	6844.8710	&	-0.00611	&	6855.5196	&	0.00208	&	6862.1313	&	-0.01179	\\
6832.9316	&	0.01076	&	6840.5653	&	0.00089	&	6848.0949	&	-0.01090	&	6855.5932	&	-0.00086	&	6862.2018	&	-0.00447	\\
6833.0026	&	0.00355	&	6840.6388	&	0.00697	&	6848.1640	&	-0.00641	&	6855.6612	&	-0.00502	&	6862.3425	&	0.00000	\\
6833.1410	&	0.00385	&	6840.7063	&	0.00911	&	6848.2380	&	-0.00618	&	6855.7340	&	0.00182	&	6862.4135	&	0.00333	\\
6833.2115	&	-0.00121	&	6840.7759	&	0.01431	&	6848.3070	&	0.00378	&	6855.8121	&	0.00233	&	6862.5539	&	0.00937	\\
6833.5631	&	-0.00313	&	6840.8476	&	0.00026	&	6848.3780	&	0.00854	&	6855.8933	&	-0.00240	&	6862.6936	&	-0.01004	\\
6833.6349	&	0.00321	&	6840.9216	&	-0.00365	&	6848.4524	&	0.01032	&	6855.9737	&	0.00021	&	6862.7658	&	-0.00697	\\
6833.7059	&	0.00287	&	6840.9972	&	-0.00386	&	6848.5216	&	0.00668	&	6856.0475	&	0.00068	&	6863.0000	&	-0.00082	\\

\hline

\end{tabular}
\end{minipage}
\end{table*}

\begin{table*}
\centering
\begin{minipage}{160mm}
\contcaption{Pulsational light curve from {\it MOST}}
\begin{tabular}{llllllllll}
\hline
JD 	&  $\triangle m$		&	JD 	&  $\triangle m$		&JD 	&  $\triangle m$		&JD 	&  $\triangle m$		&JD 	&  $\triangle m$		\\
 - 2,450,000	& (mag)	&  - 2,450,000	& (mag)	&  - 2,450,000	& (mag)	&  - 2,450,000	& (mag)	&  - 2,450,000	& (mag)	\\ \hline
6863.0706	&	0.00115	&	6870.0491	&	0.01288	&	6873.7126	&	-0.00677	&	6877.7261	&	0.00398	&	6881.5280	&	0.00519	\\
6863.1424	&	0.00144	&	6870.1210	&	0.00844	&	6873.7838	&	0.00184	&	6877.7995	&	0.00754	&	6881.5981	&	0.00608	\\
6863.2119	&	0.00268	&	6870.1926	&	0.00392	&	6873.8571	&	-0.00036	&	6877.8684	&	0.00038	&	6881.6716	&	0.00596	\\
6863.2800	&	0.00096	&	6870.2622	&	0.00049	&	6873.9283	&	0.00287	&	6877.9398	&	-0.00709	&	6881.7418	&	0.00359	\\
6863.3487	&	-0.00085	&	6870.3854	&	-0.00526	&	6874.0623	&	0.00617	&	6878.0116	&	-0.00550	&	6881.8836	&	-0.00381	\\
6863.4229	&	0.00134	&	6870.4594	&	-0.00356	&	6874.1345	&	0.00229	&	6878.0820	&	-0.00601	&	6881.9544	&	-0.00612	\\
6863.4945	&	-0.00140	&	6870.5351	&	-0.00558	&	6874.2077	&	-0.00211	&	6878.1538	&	-0.00415	&	6882.0261	&	-0.00315	\\
6863.5688	&	-0.00283	&	6870.6039	&	-0.00509	&	6874.3398	&	-0.00018	&	6878.2248	&	-0.00077	&	6882.0971	&	-0.00433	\\
6863.6381	&	-0.00315	&	6870.7437	&	0.00123	&	6874.4803	&	-0.00064	&	6878.3539	&	0.00324	&	6882.1680	&	-0.00181	\\
6863.7089	&	-0.00449	&	6870.8201	&	0.00390	&	6874.5578	&	0.00118	&	6878.4273	&	0.00431	&	6882.2406	&	0.00159	\\
6863.7811	&	0.00278	&	6870.8882	&	0.00581	&	6874.6285	&	0.00477	&	6878.4994	&	0.00827	&	6882.3702	&	0.00158	\\
6863.8473	&	0.00608	&	6870.9615	&	0.00036	&	6874.6987	&	0.00463	&	6878.5739	&	0.00705	&	6882.4418	&	0.00269	\\
6863.9221	&	0.00581	&	6871.0324	&	-0.00010	&	6874.7720	&	0.00288	&	6878.6448	&	0.00539	&	6882.5183	&	0.00185	\\
6863.9926	&	0.00099	&	6871.1037	&	-0.00104	&	6874.8405	&	0.00386	&	6878.7147	&	0.00166	&	6882.7319	&	0.00027	\\
6864.0633	&	0.00174	&	6871.1747	&	-0.00189	&	6874.9124	&	-0.00497	&	6878.7873	&	-0.00109	&	6882.8683	&	-0.00192	\\
6864.1334	&	-0.00352	&	6871.2453	&	-0.00177	&	6874.9829	&	-0.00405	&	6878.8588	&	-0.00057	&	6882.9387	&	-0.00091	\\
6864.2041	&	-0.00061	&	6871.3116	&	-0.00129	&	6875.0527	&	-0.00623	&	6878.9296	&	-0.00145	&	6883.0110	&	0.00159	\\
6864.2697	&	-0.00371	&	6871.3820	&	0.00182	&	6875.1266	&	-0.00733	&	6878.9995	&	-0.00534	&	6883.0813	&	0.00144	\\
6864.4058	&	-0.00288	&	6871.4532	&	0.00494	&	6875.1971	&	-0.00318	&	6879.0706	&	0.00181	&	6883.1515	&	0.00242	\\
6864.4808	&	0.00157	&	6871.5283	&	0.00676	&	6875.3264	&	0.00173	&	6879.1417	&	-0.00031	&	6883.2239	&	0.00110	\\
6864.5546	&	0.00161	&	6871.6011	&	0.00782	&	6875.3981	&	0.00262	&	6879.2709	&	0.00163	&	6883.3541	&	0.00053	\\
6864.6273	&	0.00324	&	6871.6711	&	0.00593	&	6875.4699	&	0.00272	&	6879.3414	&	0.00212	&	6883.4273	&	-0.00392	\\
6864.6973	&	0.00036	&	6871.7416	&	0.00096	&	6875.5433	&	0.00490	&	6879.4125	&	0.00155	&	6883.4989	&	-0.00664	\\
6864.7676	&	0.00066	&	6871.8139	&	0.00038	&	6875.6157	&	0.00125	&	6879.4843	&	-0.00358	&	6883.5767	&	-0.00564	\\
6864.8394	&	0.00152	&	6871.8843	&	-0.00202	&	6875.7549	&	-0.00044	&	6879.5587	&	-0.00395	&	6883.6489	&	-0.00184	\\
6864.9108	&	-0.00378	&	6871.9555	&	-0.00311	&	6875.8224	&	-0.00019	&	6879.6319	&	-0.00138	&	6883.7206	&	-0.00107	\\
6864.9823	&	-0.00312	&	6872.0262	&	-0.00488	&	6875.8957	&	-0.00109	&	6879.7038	&	-0.00157	&	6883.7911	&	0.00144	\\
6865.0526	&	-0.00406	&	6872.0990	&	-0.00388	&	6875.9656	&	0.00003	&	6879.7740	&	-0.00025	&	6883.9271	&	0.00721	\\
6865.1241	&	-0.00080	&	6872.1693	&	0.00232	&	6876.0364	&	0.00230	&	6879.8437	&	-0.00120	&	6884.0028	&	0.00509	\\
6865.1948	&	-0.00086	&	6872.2885	&	0.00315	&	6876.1075	&	0.00384	&	6879.9166	&	0.00305	&	6884.1373	&	0.00019	\\
6865.2653	&	0.00047	&	6872.3570	&	0.00179	&	6876.1802	&	0.00127	&	6879.9870	&	0.00503	&	6884.2097	&	-0.00112	\\
6865.3835	&	0.00258	&	6872.4298	&	0.00205	&	6876.3115	&	-0.00082	&	6880.1214	&	0.00837	&	6884.3419	&	-0.00008	\\
6865.4544	&	0.00413	&	6872.5083	&	-0.00097	&	6876.3813	&	-0.00267	&	6880.1927	&	0.00353	&	6884.4125	&	0.00185	\\
6868.7794	&	-0.00031	&	6872.5841	&	-0.00123	&	6876.4543	&	-0.00711	&	6880.2588	&	0.00112	&	6884.4841	&	-0.00080	\\
6868.8512	&	-0.00206	&	6872.6543	&	-0.00390	&	6876.5265	&	-0.00697	&	6880.3292	&	-0.00652	&	6884.5629	&	0.00767	\\
6868.9877	&	0.00180	&	6872.7254	&	-0.00454	&	6876.6017	&	-0.00727	&	6880.3996	&	-0.00576	&	6884.6335	&	0.00770	\\
6869.0585	&	0.00019	&	6872.7981	&	-0.00083	&	6876.6705	&	-0.00706	&	6880.4705	&	-0.00580	&	6884.7057	&	0.00689	\\
6869.1325	&	0.00411	&	6872.8689	&	0.00025	&	6876.7422	&	-0.00196	&	6880.5453	&	-0.00545	&	6884.7759	&	0.00727	\\
6869.2027	&	0.00472	&	6872.9405	&	0.00149	&	6876.8155	&	0.00070	&	6880.6173	&	-0.00278	&	6884.8473	&	-0.00179	\\
6869.2686	&	0.00893	&	6873.0115	&	0.00351	&	6876.8846	&	0.00901	&	6880.6868	&	-0.00145	&	6884.9189	&	0.00120	\\
6869.3385	&	0.00271	&	6873.0828	&	0.00772	&	6876.9553	&	0.01221	&	6880.7588	&	0.00431	&	6884.9894	&	-0.00493	\\
6869.4092	&	-0.00031	&	6873.2164	&	0.00183	&	6877.0278	&	0.00718	&	6880.8308	&	0.00551	&	6885.1231	&	0.00006	\\
6869.4807	&	-0.00113	&	6873.2834	&	0.00020	&	6877.1632	&	-0.00074	&	6880.9015	&	0.00287	&		&		\\
6869.5536	&	-0.00227	&	6873.3553	&	-0.01183	&	6877.2339	&	-0.00111	&	6880.9732	&	0.00176	&		&		\\
6869.6250	&	-0.00538	&	6873.4251	&	-0.01208	&	6877.3685	&	-0.00219	&	6881.0440	&	0.00207	&		&		\\
6869.6939	&	-0.00387	&	6873.4971	&	-0.01316	&	6877.5114	&	-0.00116	&	6881.1770	&	-0.00255	&		&		\\
6869.8329	&	0.00582	&	6873.5714	&	-0.01411	&	6877.5857	&	0.00066	&	6881.2480	&	-0.00396	&		&		\\
6869.9801	&	0.01055	&	6873.6427	&	-0.00979	&	6877.6559	&	0.00481	&	6881.3137	&	0.00184	&		&		\\

\hline

\end{tabular}
\end{minipage}
\end{table*}

 \begin{figure}
\includegraphics[width=80mm, angle=0]{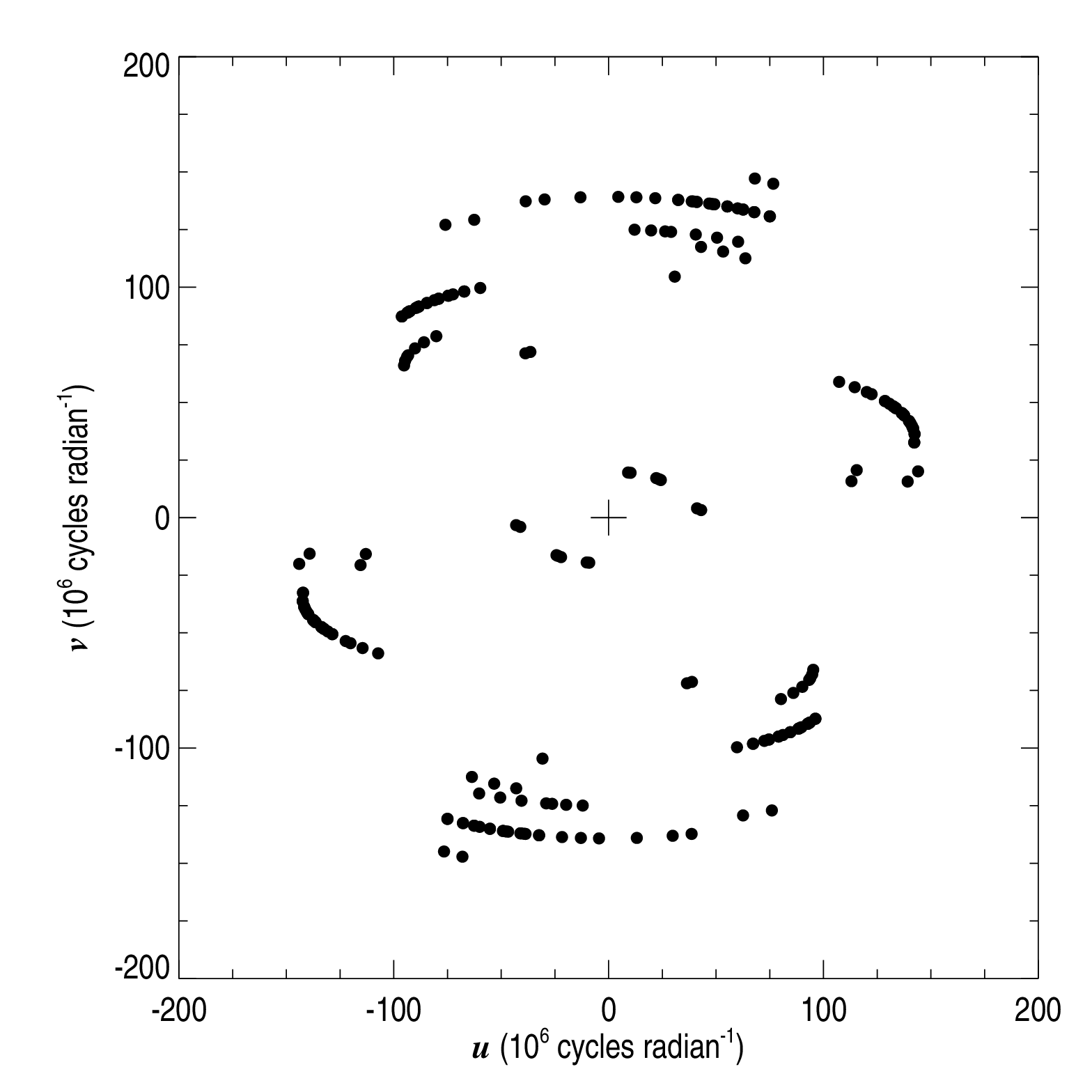}
\caption{\label{fig5} $(u,v)$ sampling of the CHARA measurements, showing that the observations sample many different baselines and position angles on the sky. }
\end{figure}

\begin{figure*}
\includegraphics[width=150mm, angle=0]{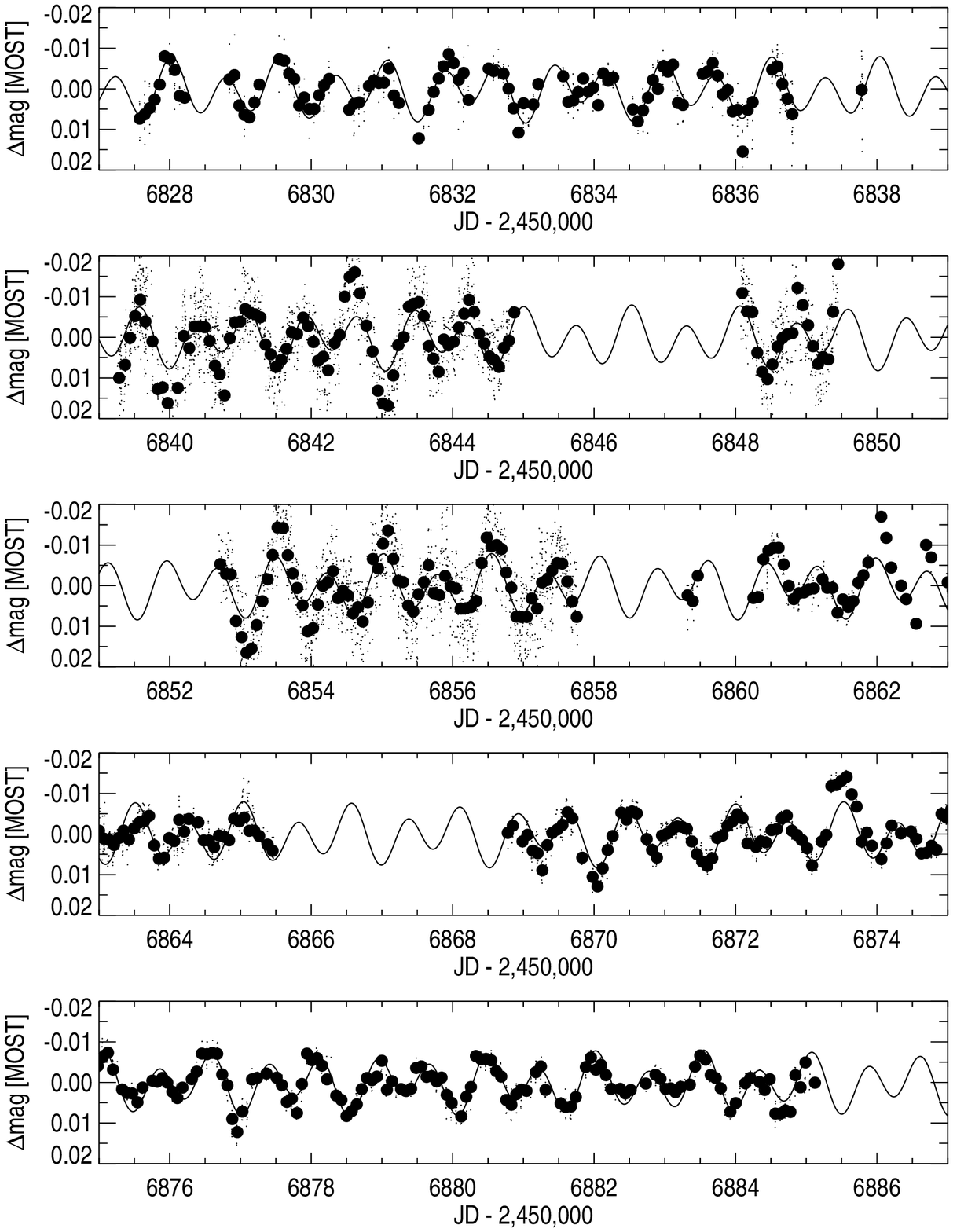}
\caption{\label{MOSTphotall} The entire de-trended \most light curve, with our two frequency fit overplotted. The detrending has removed the orbitally modulated variation, and only pulsational behaviour remains. Small points represent individual measurements, with the large points representing the orbital means.} 
\end{figure*}

\bsp \label{lastpage}

\end{document}